\documentclass[
aps,
prd,
superscriptaddress,
nofootinbib,
floatfix]
{revtex4}
\usepackage{graphicx,
longtable,
color,
bm}
\begin{document}
\title{The unfinished fabric of the three neutrino paradigm}
%
\author{        	Francesco~Capozzi}
\affiliation{   	Center for Neutrino Physics, Department of Physics, Virginia Tech, Blacksburg, VA 24061, USA}
\author{			Eleonora Di Valentino}
\affiliation{		Institute for Particle Physics Phenomenology, Department of Physics, Durham University, Durham DH1 3LE, UK}
\author{        	Eligio~Lisi}
\affiliation{   	Istituto Nazionale di Fisica Nucleare, Sezione di Bari, 
               		Via Orabona 4, 70126 Bari, Italy}
\author{        	Antonio~Marrone}
\affiliation{   	Dipartimento Interateneo di Fisica ``Michelangelo Merlin,'' 
               		Via Amendola 173, 70126 Bari, Italy}%
\affiliation{   	Istituto Nazionale di Fisica Nucleare, Sezione di Bari, 
               		Via Orabona 4, 70126 Bari, Italy}
\author{			Alessandro~Melchiorri}
\affiliation{		Dipartimento di Fisica, Universit{\`a} di Roma ``La Sapienza,'' P.le Aldo Moro 2, 00185 Rome, Italy}
\affiliation{   	Istituto Nazionale di Fisica Nucleare, Sezione di Roma~I, 
               		P.le Aldo Moro 2, 00185 Rome, Italy}
\author{        	Antonio~Palazzo}
\affiliation{   	Dipartimento Interateneo di Fisica ``Michelangelo Merlin,'' 
               		Via Amendola 173, 70126 Bari, Italy}%
\affiliation{   	Istituto Nazionale di Fisica Nucleare, Sezione di Bari, 
               		Via Orabona 4, 70126 Bari, Italy}
\begin{abstract}
\medskip
\medskip
In the current $3\nu$ paradigm, neutrino flavor oscillations probe three mixing angles $(\theta_{12},\,\theta_{23},\theta_{13})$, one CP-violating phase $\delta$, and two independent differences between the squared masses $m^2_i$, that can be chosen as $\delta m^2=m^2_2-m^2_1>0$ and $\Delta m^2=m^2_3-(m^2_1+m^2_2)/2$, where sign$(\Delta m^2)=+\,(-)$ for normal (inverted) mass ordering. Absolute $\nu$ masses can be probed by the effective mass $m_\beta$ in beta decay, by the total mass $\Sigma$ in cosmology and---if neutrinos are Majorana---by another effective mass $m_{\beta\beta}$ in neutrinoless double beta decay. Within an updated global analysis of oscillation and nonoscillation data, we constrain these $3\nu$ parameters, both separately  and in selected pairs, and highlight the concordance or discordance among different constraints. 
Five oscillation parameters $(\delta m^2,\,|\Delta m^2|,\,\theta_{12},\,\theta_{23},\theta_{13})$ are consistently measured, with an overall accuracy ranging from $\sim\! 1\%$ for $|\Delta m^2|$  to $\sim\! 6\%$ for $\sin^2\theta_{23}$ (due to its persisting octant ambiguity). We find overall hints for normal ordering (at $\sim\! 2.5\sigma$), as well as for $\theta_{23}<\pi/4$ and for  $\sin\delta<0$ (both at $90\%$ C.L.), and discuss some tensions among different datasets.  Concerning nonoscillation data, we include the recent KATRIN constraints on $m_\beta$, and we combine the latest $^{76}$Ge, $^{130}$Te and $^{136}$Xe bounds on $m_{\beta\beta}$, accounting for nuclear matrix element covariances. We also discuss some variants related to cosmic microwave background (CMB) anisotropy and lensing data, which may affect cosmological constraints on $\Sigma$ and hints on sign$(\Delta m^2)$. The default option, including all Planck results, irrespective of the so-called lensing anomaly, sets upper bounds on $\Sigma$ at the level of 
\hbox{$\sim\!10^{-1}$}~eV, and further favors normal ordering up to $\sim\! 3\sigma$.  An alternative option, that includes recent ACT results plus other independent results (from WMAP and selected Planck data) globally consistent with standard lensing, is insensitive to the ordering but prefers $\Sigma\sim \mathrm{few}\times 10^{-1}$~eV, with different  implications for $m_\beta$ and $m_{\beta\beta}$ searches. In general, the unfinished fabric of the $3\nu$ paradigm appears to be at the junction of diverse searches in particle and nuclear physics, astrophysics and cosmology, whose convergence will be crucial  to achieve a convincing completion.  
\end{abstract}
\medskip
\medskip
\maketitle

\section{Introduction}
\label{Sec:Intro}

The current three-neutrino ($3\nu$) paradigm---the simplest possibility beyond massless neutrinos 
\cite{Bilenky:2019gzn}---assumes that
the states  $(\nu_e,\,\nu_\mu,\,\nu_\tau)$ with definite flavor are mixed with three states  $(\nu_1,\,\nu_2,\,\nu_3)$ with definite masses $(m_1,\,m_2,\,m_3)$ \cite{Zyla:2020zbs}. 
In standard conventions \cite{PDG1}, the mixing matrix is parametrized by 
three angles $(\theta_{12},\,\theta_{23},\theta_{13})$ and one CP-violating phase $\delta$.
Neutrino flavor oscillations depend on such parameters and on two independent squared mass differences \cite{PDG2} that, without loss of generality, can be chosen as $\delta m^2=m^2_2-m^2_1>0$ and $\Delta m^2=m^2_3-(m^2_1+m^2_2)/2$ 
\cite{Fogli:2005cq}, with
sign$(\Delta m^2)=\pm$ distinguishing normal ordering ($+$, NO) and inverted ordering ($-$, IO). Nonoscillation 
probes of absolute neutrino masses include: $\beta$-decay, sensitive to an effective mass 
$m_\beta$ \cite{PDG3}; precision cosmology within the standard $\Lambda$CDM model \cite{PDG4}, 
sensitive to $\Sigma=m_1+m_2+m_3$ \cite{PDG5}; and, if
neutrinos are Majorana, neutrinoless double beta decay ($0\nu\beta\beta$), sensitive to another effective mass $m_{\beta\beta}$ \cite{PDG6}. 
We refer, e.g.,\ to \cite{Fogli:2005cq} for definitions of $m_\beta$ and $m_{\beta\beta}$.

Constraints on the oscillation parameters $(\delta m^2,\,\Delta m^2,\,\theta_{ij},\,\delta)$ and on the nonoscillation observables $(\Sigma,\,m_\beta,\,m_{\beta\beta})$ have been explored in several global neutrino data analyses, 
including our previous work \cite{Capozzi:2017ipn} and the more recent papers \cite{Capozzi:2020,deSalas:2020pgw,Esteban:2020cvm}, plus the preliminary contribution in \cite{Marrone2021}. 
In particular, the analyses in \cite{deSalas:2020pgw,Esteban:2020cvm,Marrone2021} are based on largely common datasets, 
based on updated information from the Conference {\em Neutrino 2020\/} \cite{Nu2020}. As a result, 
a solid fabric for the $3\nu$ paradigm emerges from convergent measurements of 
five oscillation parameters $(\theta_{12},\,\theta_{23},\theta_{13},\,\delta m^2,\,|\Delta m^2|)$, with an overall accuracy ranging from $\sim\! 1\%$ for $|\Delta m^2|$  to $\sim\! 6\%$ for $\sin^2\theta_{23}$ (dominated by the so-called $\theta_{23}$ octant degeneracy). However, the fabric is still unfinished in $\nu$ oscillations, as far as the $\theta_{23}$ octant, 
the mass ordering and the phase $\delta$ are concerned. 
In particular, a tension between recent long-baseline accelerator neutrino data (from T2K \cite{T2K2020} and NOvA \cite{NOvA2020}) 
affects all these $3\nu$ unknowns at the same time \cite{deSalas:2020pgw,Esteban:2020cvm,Marrone2021,Kelly:2020fkv}.
In addition, the Dirac-Majorana nature
and the absolute neutrino mass scale remain undetermined in current nonoscillation searches, with $\Sigma$, $m_\beta$ and $m_{\beta\beta}$ constrained at sub-eV scales but still consistent with null values \cite{Formaggio:2021nfz}.

In this work we discuss the status of the $3\nu$ framework, including new data that have recently become available
from both oscillation and nonoscillation searches, with particular attention to issues of concordance or discordance of various data sets, and to their implications on the unfinished fabric of the paradigm. By performing a global analysis of oscillation data, including the latest Super-Kamiokande atmospheric results made publicly available in 2021 \cite{SKmap}, 
we find an indication for normal ordering at the level of $2.5\sigma$, as well as $90\%$ C.L.\ hints for $\theta_{23}<\pi/4$ and for  $\sin\delta<0$.  We discuss the structure and interplay of such hints, especially in the light of the T2K and NOvA tension and of the complementarity among accelerator, atmospheric and reactor data. We surmise that further understanding of neutrino nuclear interactions may help to clarify some issues. 

Concerning nonoscillation searches, we include the KATRIN 2021 data \cite{KATRIN2021} that, for the first time, set sub-eV upper bounds on $m_\beta$ at 90\% C.L. We analyze systematically all the latest $0\nu\beta\beta$ decay searches probing half lives $T>10^{25}$~y (in $^{76}$Ge, $^{130}$Te and $^{136}$Xe), and translate them into $m_{\beta\beta}$ bounds via correlated nuclear matrix elements. In the realm of cosmology and of its consensus $\Lambda$CDM model, increasing attention is also being paid to old and new data tensions, see e.g.\ \cite{DiValentino:2021izs,DiValentino:Snowmass,Challenge2021}. In this context, we focus on the 
so-called $A_\mathrm{lens}$ anomaly affecting Planck angular spectra (that show more lensing
than expected in the $\Lambda$CDM model \cite{Aghanim:2018eyx}), and we consider two possible  
options, leading to different implications for absolute mass observables. On the one hand, we revisit a previously considered ``default'' scenario \cite{Capozzi:2020},
including all Planck results (irrespective of the $A_\mathrm{lens}$ anomaly), that sets stringent upper bounds on $\Sigma$ at the level of $\sim\! 10^{-1}$~eV, and further favors normal ordering, raising its overall preference to $\sim 3\sigma$. On the other hand, we discuss an ``alternative'' option
that makes use of the recent ACT CMB polarization data release 4 (ACTPol-DR4)  \cite{Aiola:2020azj} that is consistent with standard lensing, also in combination
with WMAP 9-year data (WMAP9) \cite{Bennett:2012zja} and selected data from Planck \cite{Aghanim:2018eyx,Aghanim:2019ame,Aghanim:2018oex}; such option is insensitive to the mass ordering and prefers $\Sigma\sim \mathrm{few}\times 10^{-1}$~eV, with different  implications for $m_\beta$ and $m_{\beta\beta}$ searches.

Building upon our previous works \cite{Capozzi:2017ipn,Capozzi:2020} we elaborate upon these recent topics as follows: 
In Sections~II and III we update and discuss the analysis of oscillation and nonoscillation data, respectively.
We pay particular attention to relevant correlations among various observables, and to some emerging tensions 
among different data sets. We provide a brief synthesis of oscillation and nonoscillation results in Sec.~IV.

\vspace*{-1mm}
\section{Oscillation data, analysis methods and results}
\label{Sec:Osc}

In this section we introduce recent oscillation data that were not included in our previous work \cite{Capozzi:2020}, 
together with the methodology used for their analysis, in the light of some emerging issues in precision oscillation physics. We then discuss the resulting constraints on the parameters $(\delta m^2,\,\Delta m^2,\,\sin^2\theta_{ij},\,\delta)$,
both separately and in selected pairs, highlighting the concordance, discordance and complementarity of various datasets.

\vspace*{-1mm}
\subsection{Oscillation data update}

Three-neutrino oscillations are currently constrained by experiments using long-baseline (LBL) accelerator, solar, long-baseline reactor (KamLAND), short-baseline (SBL) reactor and atmospheric neutrinos. With respect to \cite{Capozzi:2020}, we update some of these datasets as follows. LBL accelerator data, in the form of neutrino and antineutrino energy spectra for flavor disappearance and appearance channels, are taken from the presentations of T2K \cite{T2K2020} and NOvA \cite{NOvA2020} at {\em Neutrino 2020\/} \cite{Nu2020} (and subsequent conferences). Such spectra are endowed with statistical (Poisson) errors and 
systematic (normalization and energy scale) uncertainties, as well as with oscillation-independent backgrounds, 
in a modified version of GLOBES \cite{GLOBES}. Solar neutrinos data from the Super-Kamiokande-IV 2970-days run (SK-IV  energy spectrum and day-night asymmetry) are taken from the presentation at {\em Neutrino 2020\/} \cite{SK2020}, while the input solar model BP16-GS98 \cite{Vinyoles:2016djt} is unchanged.  Concerning SBL reactors, we update from {\em Neutrino 2020\/} the RENO \cite{RENO2020} and 
Double Chooz data \cite{DC2020}, while the
Daya Bay data \cite{Adey:2018zwh} are unchanged. Note that Daya Bay and and RENO measure both 
$\theta_{13}$ and $\Delta m^2$, while the latter parameter is not significantly constrained by
Double Chooz. IceCube-DeepCore (IC-DC) atmospheric data 
are taken as in \cite{Capozzi:2020}; a new IC-DC data release  is expected in the near future \cite{ICDC2021}.
Finally, our oscillation dataset is completed by the recent SK-IV atmospheric results \cite{SK2020,Jiang:2019xwn}, included through the
$\chi^2$ map  recently made available by the collaboration \cite{SKmap}.

\vspace*{-1mm}
\subsection{Analysis method and emerging issues}
 
We adopt the methodology proposed in \cite{Capozzi:2018ubv}, see also  \cite{Capozzi:2017ipn,Capozzi:2020}. In particular, 
we start with the combination of solar, KamLAND and LBL accelerator neutrino data, that represent 
the minimal dataset sensitive to all the oscillation parameters $(\delta m^2,\,\Delta m^2,\,\theta_{ij},\,\delta)$.
We then add SBL reactor neutrino data, that
sharpen the constraints on $(|\Delta m^2|,\theta_{13})$ and indirectly affect the parameters $(\theta_{23},\,\delta)$
and sign$(\Delta m^2)$ via correlations. We add atmospheric neutrino data at the end, for two reasons: (a) they
 provide rich but
rather entangled information on the parameters $(\Delta m^2,\theta_{23},\,\theta_{13},\,\delta)$; (b) 
their $\chi^2(\Delta m^2,\theta_{23},\,\delta)$ maps
assume an input on $(\delta m^2,\,\theta_{12},\,\theta_{13})$ from the combination of solar, KamLAND and SBL reactor data.
Finally, a frequentist approach based on $\chi^2$ functions is used for all datasets. Best fits
are obtained by $\chi^2$ minimization, while allowed regions around best fits are expanded  
in terms of ``number of standard deviations'' $N_\sigma=\sqrt{\Delta \chi^2}$. 
\textcolor{black}{In particular, two-dimensional contours
are shown for $N_\sigma=1$, 2 and 3, which, for a $\chi^2$ distribution with two degrees of freedom, correspond
to C.L.\ of 39.35\%, 86.47\% and 98.89\%, respectively. Their one-dimensional 
projections provide the $N_\sigma$ ranges for each parameter, corresponding to C.L.\ 
of  68.27\%, 95.45\% and 99.73\%, respectively.}
The difference $\Delta \chi^2_{\mathrm{IO}-\mathrm{NO}}$ between the minima in IO and NO may---or may not---be accounted
for, when reporting fit results; these two options will be clearly distinguished in each context. 

We briefly discuss some issues arising in global data analyses, in 
the era of increasingly precise measurements and of growing sensitivity to subleading effects. 
Data fits  usually
involve the comparison of experimental event rates $R_\mathrm{expt}$ 
with their theoretical predictions $R_\mathrm{theo}$
\begin{equation}
\label{Rtheo} 
R_\mathrm{theo} = \int \Phi_\alpha \otimes P_{\alpha\beta} \otimes \sigma_\beta \otimes r_\beta \otimes  \varepsilon_\beta\ ,
\end{equation}
where, from left to right, the integrands represent the source flux of $\nu_\alpha$, the probability of
$\nu_\alpha\to\nu_\beta$ oscillations, and the interaction cross section,  detector resolution and efficiency for 
$\nu_\beta$ events. Some 
factors may be  differential functions that need multiple integrations or convolutions, 
as alluded by the cross product $\otimes$. Integrands are endowed with various uncertainties, that may be shared
by (i.e., correlated among) various rates $R$ in the same or different experiment(s).
In solar neutrino searches, all these features can be accurately implemented to a large extent
\cite{Fogli:2002pt}. Also short-baseline reactor experiments (Daya Bay, RENO, Double Chooz) 
generally provide enough public information to allow reproducible analyses, although 
a more precise joint analysis by the different collaborations (accounting for minor correlated uncertainties)
would be desiderable \cite{ESCAPE}. 
More relevant issues arise in the context of long-baseline accelerator searches, currently carried out
by T2K \cite{T2K2020} and NOvA \cite{NOvA2020}. 
Their event spectra are usually given in terms of a ``reconstructed'' (unobservable) neutrino energy $E_\nu^\mathrm{rec}$, that
is processed from observable event energies at far detectors, through models (for some $R_\mathrm{theo}$ integrands)  
constrained by near-detector data. 
In principle, both T2K and NOvA should share a common theoretical model for $\sigma$ (and to some extent for $\Phi$), 
leading to possible correlations among their uncertainties for $E_\nu^\mathrm{rec}$ (affecting $\Delta m^2$) and for
the event
rates $R_\mathrm{theo}(E_\nu^\mathrm{rec})$ (affecting $\theta_{23}$ and $\theta_{13}$). In practice, the adopted models are different, and possible covariances are ignored. A joint analysis planned
by the T2K and NOvA 
collaborations \cite{JOINT} might shed light on these issues.
Finally, in current atmospheric neutrino searches at SK and IC-DC, 
the data processing and analysis are too complex to be reproducible by external users with an acceptable accuracy. Oscillation results are
given in terms of public  $\chi^2$ maps that, when summed up, cannot account for known covariances, 
such as those related to the (common, in principle) input models for $\Phi$ and $\sigma$. Once again, joint 
analyses or in-depth comparisons of data by different atmospheric $\nu$ experiments would be desirable \cite{ATMOS}. We surmise that
global oscillation analyses could 
benefit from a better control of those systematics that are shared by different experiments 
(such as model uncertainties for $\Phi_\alpha$ and $\sigma_\beta$), but whose correlations are not yet
properly accounted for. See also the remarks in Sec.~\ref{sec:remarks}.
  
\vspace*{-1mm}
\subsection{Results on single oscillation parameters}

In this Section we present the constraints on the six oscillation parameters $(\delta m^2,\,\Delta m^2,\,\sin^2\theta_{ij},\,\delta)$ for increasingly rich data sets. We explicitly account for the $\chi^2$ difference between NO and IO, in order to show its variations. 

\begin{figure}[t!]
\begin{minipage}[c]{0.85\textwidth}
\includegraphics[width=0.82\textwidth]{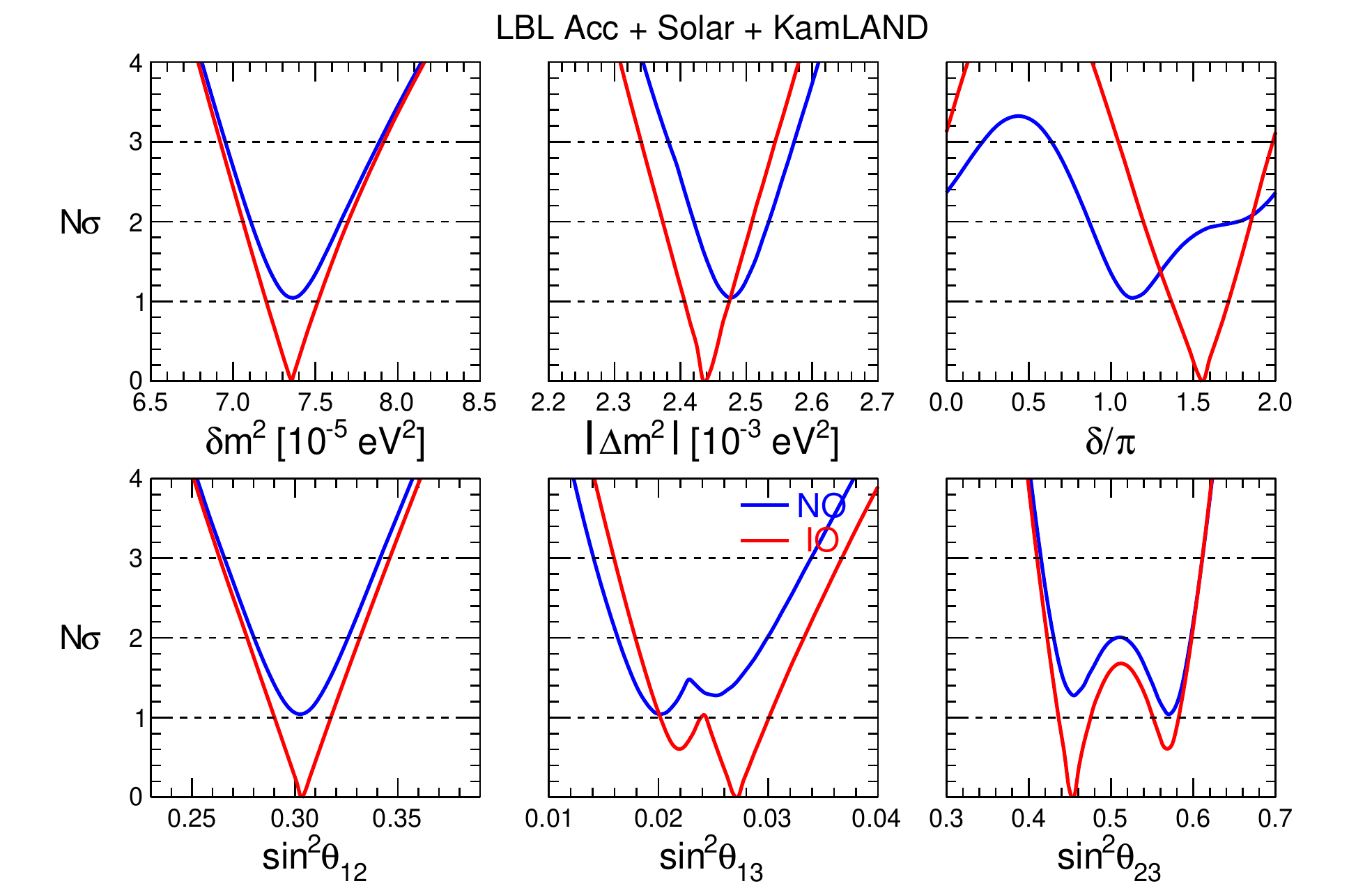}
\caption{\label{Fig_01}
\footnotesize Global $3\nu$ oscillation analysis of long-baseline accelerator, solar and KamLAND $\nu$  data. 
Bounds on the parameters $\delta m^2$, $|\Delta m^2|$, $\sin^2\theta_{ij}$, and $\delta$, for NO (blue) and IO (red), in terms of $N_\sigma=\sqrt{\Delta \chi^2}$ from the global best fit. The offset between separate minima in IO and  NO, 
$\Delta \chi^2_\mathrm{IO-NO}=-1.1$, favors the IO case by $\sim\! 1.0\sigma$.   
} \end{minipage}
\end{figure}

Figure~\ref{Fig_01} shows the results for the combination of solar and KamLAND data 
(sensitive to $\delta m^2$, $\sin^2\theta_{12}$, and $\sin^2\theta_{13}$) with LBL accelerator data 
(mainly sensitive to $\Delta m^2$, $\sin^2\theta_{23}$, $\sin^2\theta_{13}$ and $\delta$), for
both NO (blue) and IO (red). The latter mass ordering is slightly favored (by accelerator data) at the level of $\sim\!1\sigma$, as also discussed later.
The parameters $\delta m^2$ and $\sin^2\theta_{12}$ are rather precisely measured, with nearly linear and symmetrical 
(i.e., almost gaussian) uncertainties, and no significant difference between constraints in NO and IO.
The parameters $\sin^2\theta_{13}$ and $\sin^2\theta_{23}$ are less 
accurately constrained. In particular, the two minima in $\sin^2\theta_{23}$ reflect the $\theta_{23}$ 
octant ambiguity in the $\nu_\mu\to\nu_\mu$ disappearance searches at LBL accelerators, 
inducing two correlated minima in  $\sin^2\theta_{13}$ via the leading amplitude
of $\nu_\mu\to\nu_e$ appearance ($\propto \sin^2\theta_{23}\sin^2\theta_{13}$) \cite{Fogli:1996pv}.
The phase $\delta$ is poorly constrained, although it appears to be 
slightly favored around $\pi$ in NO and around $3\pi/2$ in IO, while it is disfavored 
around $\pi/2$ in both cases.   

\begin{figure}[h!]
\begin{minipage}[c]{0.85\textwidth}
\includegraphics[width=0.82\textwidth]{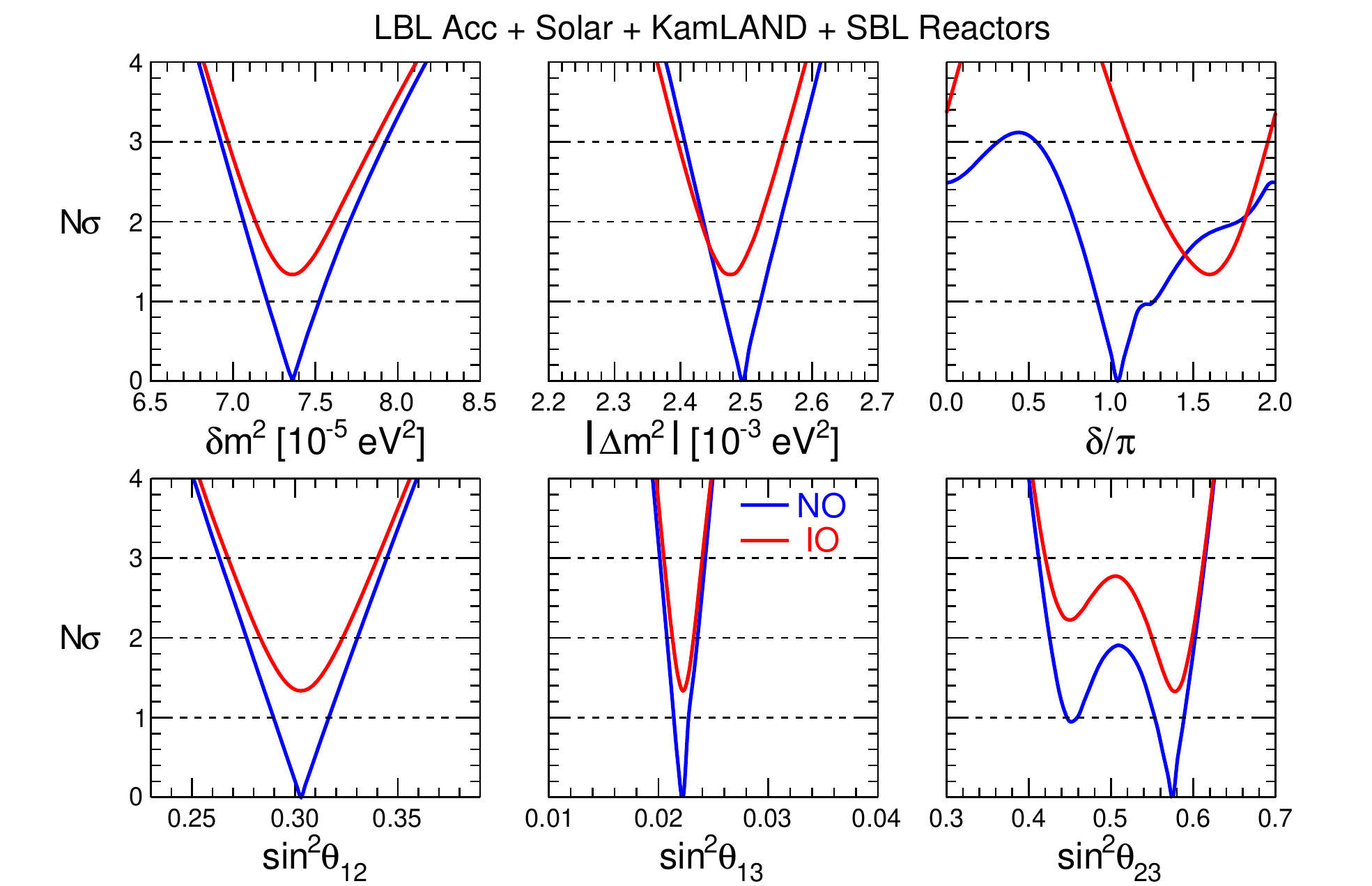}
\caption{\label{Fig_02}
\footnotesize As in Fig.~\protect\ref{Fig_01}, but adding short-baseline reactor $\nu$ data.   
The offset $\Delta \chi^2_\mathrm{IO-NO}=+1.8$  favors the NO case by $\sim\!1.3\sigma$.   
} \end{minipage}
\end{figure}

Figure~\ref{Fig_02} shows the effect of adding SBL reactor data, which are sensitive to 
$|\Delta m^2|$ and $\sin^2\theta_{13}$. One can notice the strong reduction of the 
$\sin^2\theta_{13}$ uncertainty, inducing also correlated changes on the relative likelihood 
of the lower and upper octant of $\theta_{23}$ via $\nu_\mu\to\nu_e$ appearance in LBL accelerators
\cite{Fogli:1996pv}. The synergy of SBL reactor and LBL accelerator data 
also helps to break mass-ordering degeneracies via independent measurements of $\Delta m^2$
(see, e.g.,   \cite{Huber:2003pm}) and currently flips the fit preference from IO to NO (at 
the level of $\sim\!1.3\sigma$), together with an increase of the best-fit value of $\Delta m^2$
with respect to Fig.~\ref{Fig_01}. The preference for $\delta\sim \pi$ ($\sim 3\pi/2$) in NO (IO) remains
unaltered.


\vspace*{10mm}

\newpage

\begin{figure}[t!]
\begin{minipage}[c]{0.85\textwidth}
\includegraphics[width=0.82\textwidth]{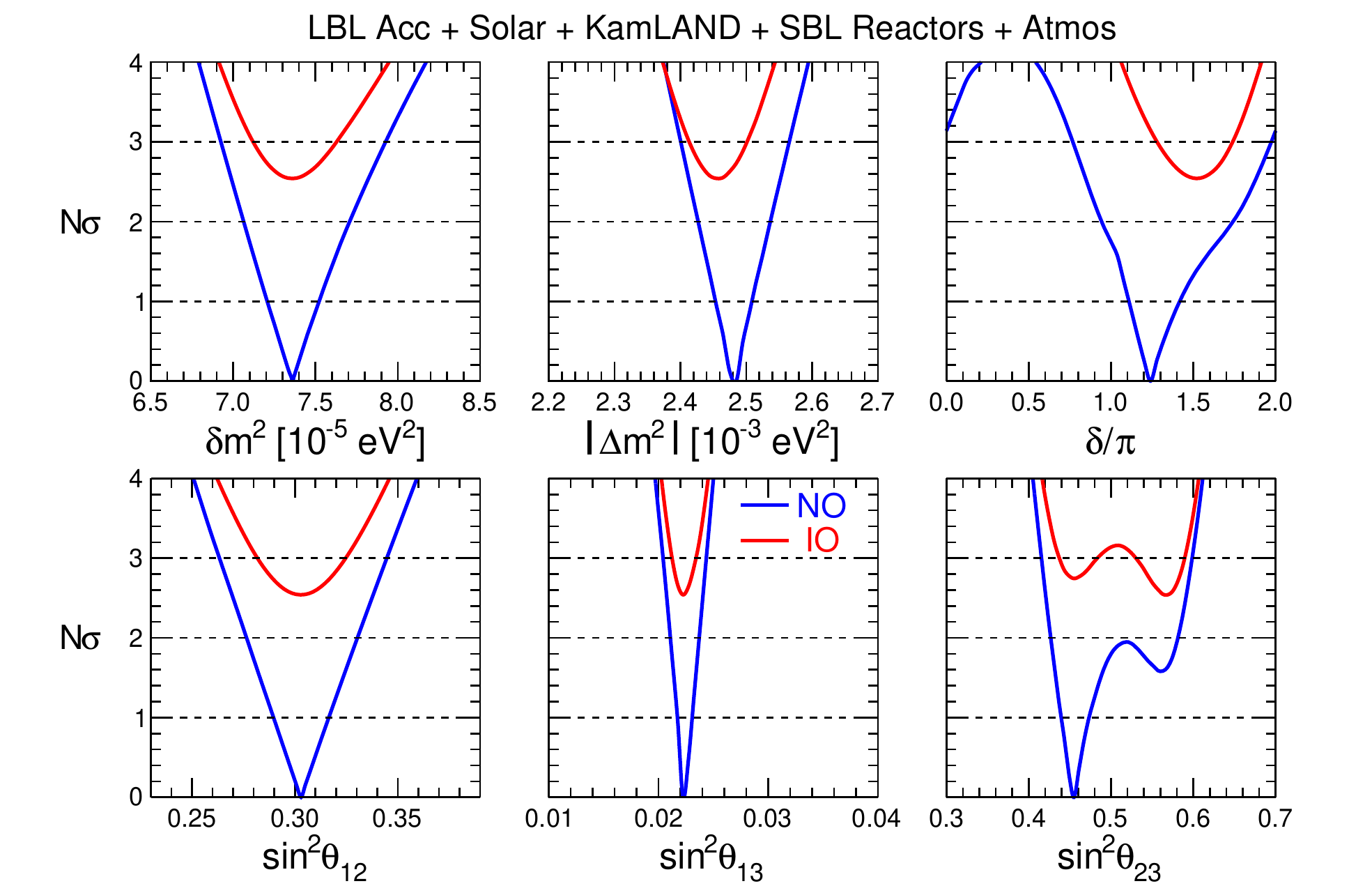}
\caption{\label{Fig_03}
\footnotesize As in Fig.~\protect\ref{Fig_02}, but adding atmospheric $\nu$ data (i.e., with
all oscillation data included).   
The offset $\Delta \chi^2_\mathrm{IO-NO}=+6.5$ favors the NO case by $\sim\!2.5\sigma$.   
} 
\end{minipage}
\end{figure}

Figure~\ref{Fig_03} shows the effect of adding atmospheric $\nu$ data, which add further sensitivity
to $\Delta m^2$ (and to its sign), as well as to $\sin^2\theta_{23}$ and $\delta$. In particular, the inclusion of SK-IV 
data  \cite{SK2020,SKmap} corroborates the preference in favor of NO (at an overall level of $\sim\!2.5\sigma$),
flips the $\theta_{23}$ preference from the upper to the lower octant in NO (at $\sim 1.6\sigma$) 
and also moves the best fit of $\delta$ slightly above the CP-conserving value $\pi$ (disfavored at $\sim 1.6\sigma$). 
The latter hints in favor of  $\theta_{23}<\pi/4$  and on $\delta> \pi$, 
currently emerging in NO at the statistical ``threshold of interest''  
of 90\% C.L., represent interesting updates with respect to previous global analyses not including SK-IV
atmospheric data  \cite{deSalas:2020pgw,Esteban:2020cvm,Marrone2021}. 

Table~\ref{Tab:Synopsis} reports a numerical summary of the same information shown in Fig.~3, for 
the separate cases of NO and IO (whose $\chi^2$ difference is reminded in the last row). 
The two squared mass splittings $\Delta m^2$ and $\delta m^2$ are measured with a formal $1\sigma$ accuracy
of $1.1\%$ and $2.3\%$, respectively. The mixing parameters $\sin^2\theta_{13}$, $\sin^2\theta_{12}$ and $\sin^2\theta_{23}$
are measured with an accuracy of $\sim\! 3\%$, $4.5\%$, and $\sim 6\%$, respectively.
The latter uncertainty is largely affected by the $\theta_{23}$ octant ambiguity; if one of the two quasi-degenerate $\theta_{23}$ options could be removed, such uncertainty would be reduced by factor of $\sim\!2$ in both NO and IO.

Summarizing, five oscillation parameters are known with (few) percent accuracy, while 
only some hints emerge about the remaining three oscillation ``unknowns''. In particular, 
we find a preference for NO at $\sim\!2.5\sigma$ and, in such ordering, we also find a preference at 90\% C.L.\ for  
$\theta_{23}$ in the lower octant (with respect to the secondary best fit
 in the upper octant) and for $\delta\simeq 1.24\pi$ (with respect to the CP-conserving value $\delta=\pi$). Conversely,
maximal $\theta_{23}$ mixing is disfavored at $\sim 1.8\sigma$ and  
the range $\delta \in [0,\, 0.77\pi]$ is disfavored at $>3\sigma$ in NO.

\begin{table}[h!]
\centering
\resizebox{.82\textwidth}{!}{\begin{minipage}{\textwidth}
\caption{\label{Tab:Synopsis} 
Global $3\nu$ analysis of oscillation parameters: best-fit values and allowed ranges at $N_\sigma=1$, 2 and 3, for  either NO or  IO, including all data. The latter column shows the formal  ``$1\sigma$ fractional accuracy'' for each parameter, defined as 1/6 of the $3\sigma$ range, divided by the best-fit value and expressed in percent. We recall that 
$\Delta m^2=m^2_3-{(m^2_1+m^2_2})/2$ and that $\delta \in [0,\,2\pi]$ (cyclic). The last row reports the  difference
between the $\chi^2$ minima in IO and NO.
}
\begin{ruledtabular}
\begin{tabular}{lcccccc}
Parameter & Ordering & Best fit & $1\sigma$ range & $2\sigma$ range & $3\sigma$ range & ``$1\sigma$'' (\%) \\
\hline
$\delta m^2/10^{-5}~\mathrm{eV}^2 $ & NO, IO & 7.36 & 7.21 -- 7.52 & 7.06 -- 7.71 & 6.93 -- 7.93 & 2.3 \\
\hline
$\sin^2 \theta_{12}/10^{-1}$ & NO, IO & 3.03 & 2.90 -- 3.16 & 2.77 -- 3.30 & 2.63 -- 3.45 & 4.5 \\
\hline
$|\Delta m^2|/10^{-3}~\mathrm{eV}^2 $ & NO  & 2.485 & 2.454 -- 2.508 & 2.427 -- 2.537 & 2.401 -- 2.565 & 1.1 \\
                                      & IO  & 2.455 & 2.430 -- 2.485 & 2.403 -- 2.513 & 2.376 -- 2.541 & 1.1 \\
\hline
$\sin^2 \theta_{13}/10^{-2}$ & NO & 2.23 & 2.17 -- 2.30 & 2.11 -- 2.37 & 2.04 -- 2.44 & 3.0 \\
                             & IO & 2.23 & 2.17 -- 2.29 & 2.10 -- 2.38 & 2.03 -- 2.45 & 3.1 \\
\hline
$\sin^2 \theta_{23}/10^{-1}$ & NO & 4.55 & 4.40 -- 4.73 & 4.27 -- 5.81 & 4.16 -- 5.99 & 6.7 \\
                             & IO & 5.69 & 5.48 -- 5.82 & 4.30 -- 5.94 & 4.17 -- 6.06 & 5.5 \\
\hline
$\delta/\pi$ & NO & 1.24 & 1.11 -- 1.42 & 0.94 -- 1.74  &  0.77 -- 1.97  & 16 \\
             & IO & 1.52 & 1.37 -- 1.66 & 1.22 -- 1.78  &  1.07 -- 1.90  & 9 \\
\hline
$\Delta \chi^2_{\mathrm{{IO}-{NO}}}$ & IO$-$NO & +6.5  \\ [1pt]
\end{tabular}
\end{ruledtabular}
\end{minipage}}
\end{table}

\begin{figure}[t]
\begin{minipage}[c]{0.85\textwidth}
\includegraphics[width=0.34\textwidth]{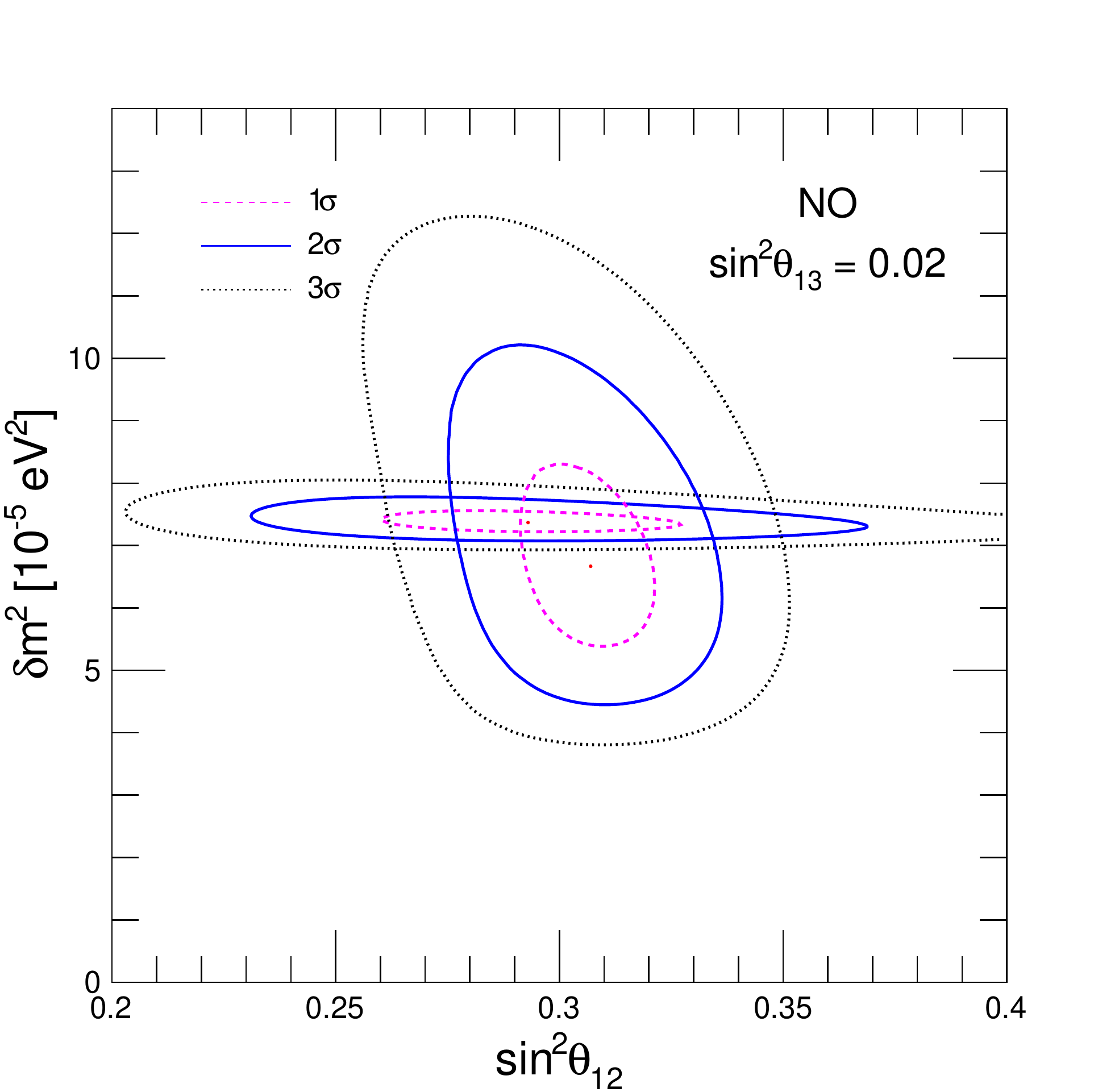}
\vspace*{-2.5mm}
\caption{\label{Fig_04}
\footnotesize 
Regions separately allowed by solar and KamLAND data in the plane $(\sin^2\theta_{12},\,\delta m^2)$ for  
$\sin^2\theta_{13}=0.02$ and NO. (The case of IO, not shown, would be almost identical). 
The solar $\nu$ fit includes SK-IV 2970-day data \cite{SK2020}.}
\end{minipage}
\vspace*{-3mm}
\end{figure}

\subsection{Results on selected pairs of oscillation variables}
\vspace*{-2mm}

By studying selected pairs of variables  we can gain 
further insights about current unknowns (the mass ordering, the octant of $\theta_{23}$ and the CP phase $\delta$), 
and appreciate their interplay with known features of $3\nu$ oscillations. We discuss the pairs 
 $(\sin^2\theta_{12},\,\delta m^2)$, $(\sin^2\theta_{23},\,\sin^2\theta_{13})$, 
\textcolor{black}{$(\sin^2\theta_{23},\,|\Delta m^2|)$}, 
$(\sin^2\theta_{23},\,\delta)$, as well as pairs of total 
$\nu_e$ and $\overline\nu_e$ events (bi-event plots) as observed in the appearance channel by T2K and NOvA.

Figure~\ref{Fig_04} shows the regions separately allowed by solar and KamLAND neutrino data 
in the plane charted by $(\sin^2\theta_{12},\,\delta m^2)$, assuming  fixed $\sin^2\theta_{13}=0.02$ and NO.
The two regions were somewhat displaced in the past, leading to a $<2\sigma$ tension 
between the best-fit $\delta m^2$ values \cite{PDG1} (see, e.g., the analogous Fig.~4 in \cite{Capozzi:2018ubv}). The 
current regions in Fig.~\ref{Fig_04} appear to be in very good agreement, largely as a 
result of a slightly smaller day-night asymmetry in SK-IV 2970-day solar data, shifting
the solar $\delta m^2$ best fit upwards and closer to the KamLAND one \cite{SK2020}. 
We find that this shift does not alter the combined solar and KamLAND constraints on 
$\theta_{13}$, namely, $\sin^2\theta_{13}\simeq 0.014\pm 0.015$ 
 (see, e.g., Fig.~5 in \cite{Capozzi:2018ubv}). Results for IO (not shown)
would be almost identical for all parameters $(\delta m^2,\,\sin^2\theta_{12},\,\sin^2\theta_{13})$.
In conclusion, solar and KamLAND data are
not only in very good agreement about the $(\nu_1,\,\nu_2)$ oscillation parameters
$(\delta m^2,\,\sin^2\theta_{12})$, but are also consistent with the 
measurement $\sin^2\theta_{13}\simeq 0.02$ at SBL reactors.

\begin{figure}[b]
\begin{minipage}[c]{0.85\textwidth}
\includegraphics[width=0.6\textwidth]{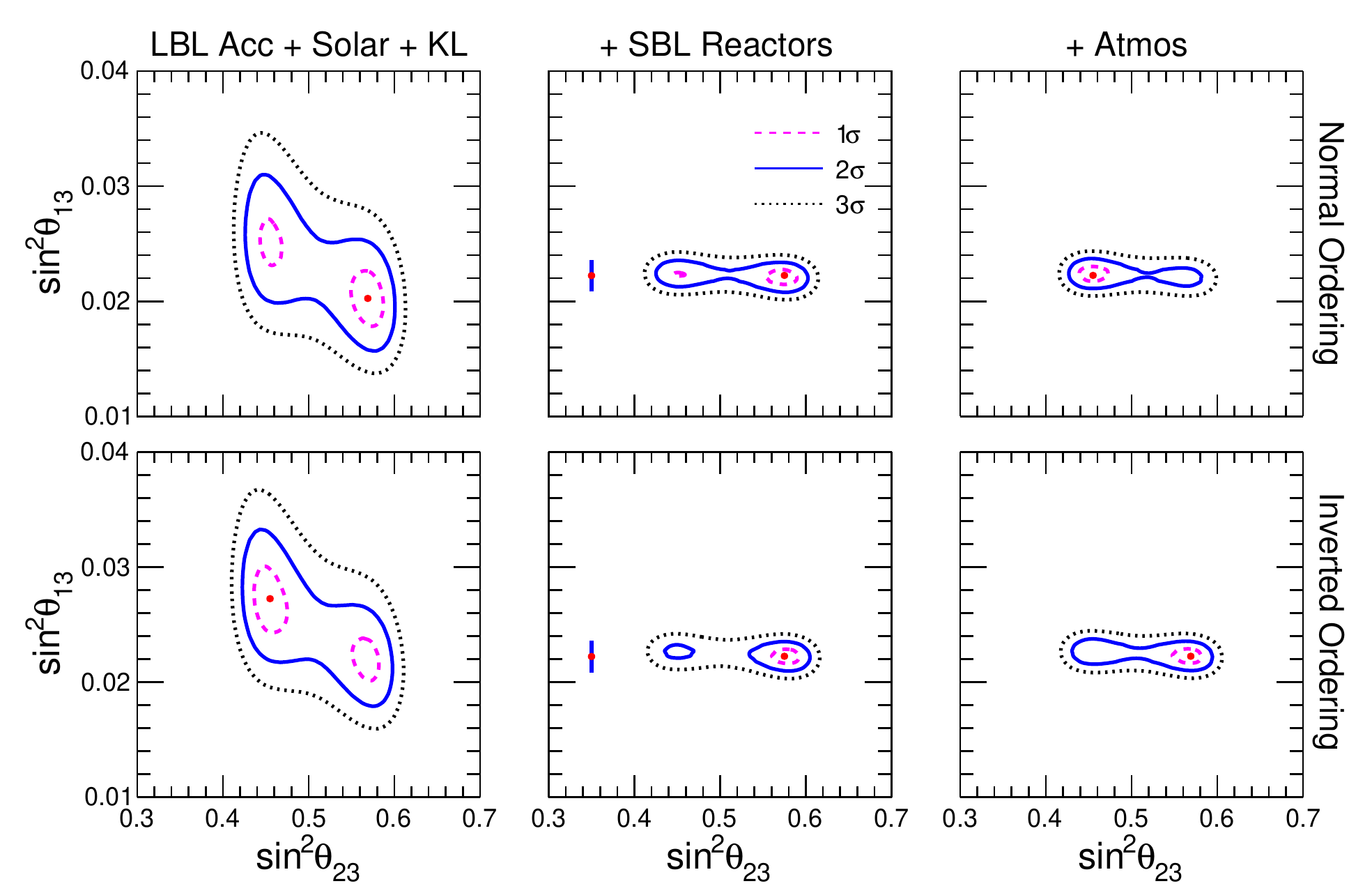}
\vspace*{-2.5mm}
\caption{\label{Fig_05}
\footnotesize Regions allowed in the plane $(\sin^2\theta_{23},\,\sin^2\theta_{13})$ for increasingly rich data sets:
Solar + KamLAND + LBL accelerator data (left panels), plus SBL reactor data (middle panels), plus Atmospheric data (right panels). Top and bottom panels refer, respectively, to NO and IO as taken separately (i.e., without any relative $\Delta\chi^2$ offset). The error bars in the middle panels show the $\pm2\sigma$ range for $\theta_{13}$ arising from SBL reactor data only. 
}
\end{minipage}
\end{figure}

Figure~\ref{Fig_05} shows the covariance of the pair $(\sin^2\theta_{23},\,\sin^2\theta_{13})$
for increasingly rich data sets, in both NO (top) and IO (bottom), 
\textcolor{black}{with the corresponding $\chi^2$ functions separately minimized for each mass ordering}. 
The $\theta_{23}$ octant ambiguity leads to two quasi-degenerate solutions at $1\sigma$, that generally merge 
at $\sim \!2\sigma$.  The leading appearance amplitude in LBL accelerators, scaling as $\sin^2\theta_{23}\sin^2\theta_{13}$, 
induces an anticorrelation between the two angles in the left panels: the higher $\theta_{23}$, the smaller $\theta_{13}$.
In the middle panels, the results from SBL reactors only (represented by $\pm 2\sigma$ error bars) tend to prefer
slightly  the upper-octant solution
(with lower values of $\theta_{13}$) in both NO and IO, 
as confirmed by the combination with LBL accelerators (continuous curves). In the right panels,
however, adding atmospheric data (that include SK-IV \cite{SK2020,SKmap}) flips the octant preference in NO, while
confirming it in IO. We conclude that current hints about
the $\theta_{23}$ octant are still rather fragile. 
\vspace*{-4mm}

\newpage

\begin{figure}[t!]
\begin{minipage}[c]{0.85\textwidth}
\includegraphics[width=0.6\textwidth]{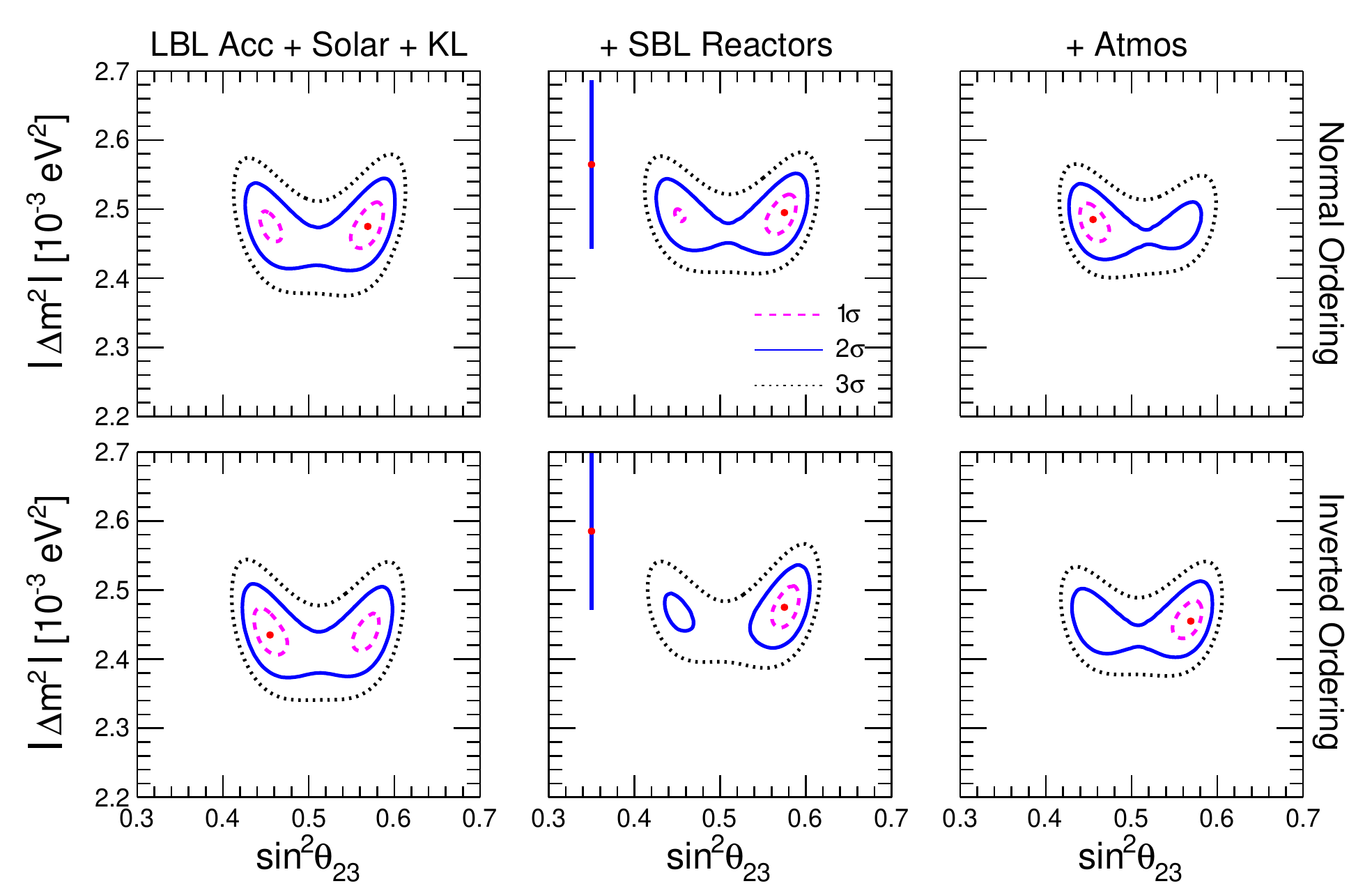}
\vspace*{-2mm}
\caption{\label{Fig_06}
\footnotesize As in Fig.~\ref{Fig_05}, but in the plane 
\textcolor{black}{$(\sin^2\theta_{23},\,|\Delta m^2|)$.}
 The error bars in the middle panels show the $\pm2\sigma$ range for 
 \textcolor{black}{$|\Delta m^2|$} 
 arising from SBL reactor data only. 
}
\end{minipage}
\end{figure}

\begin{figure}[b!]
\begin{minipage}[c]{0.85\textwidth}
\includegraphics[width=0.6\textwidth]{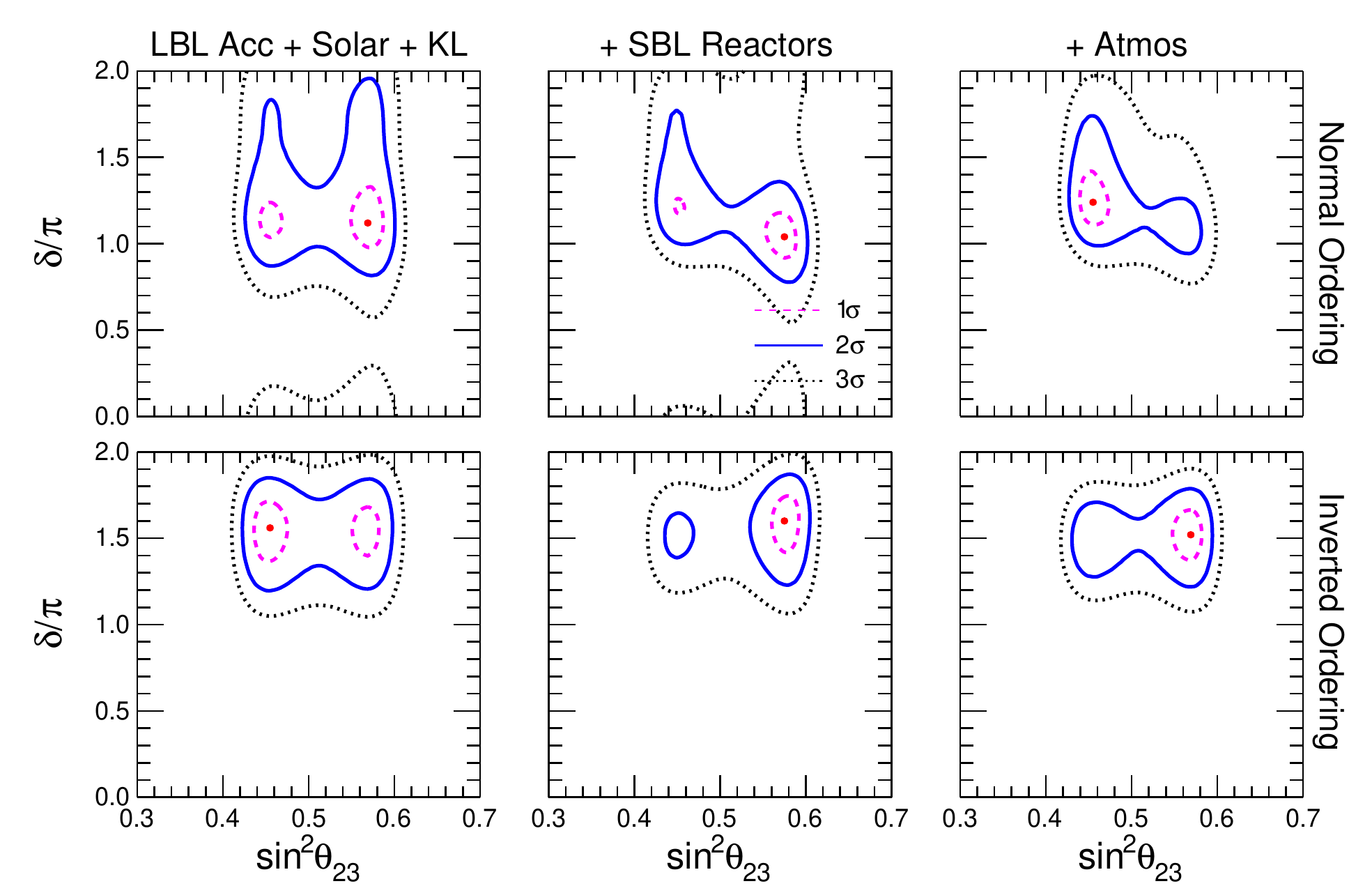}
\vspace*{-2mm}
\caption{\label{Fig_07}
\footnotesize As in Fig.~\ref{Fig_06}, but in the plane $(\sin^2\theta_{23},\,\delta)$.
}
\end{minipage}
\end{figure}

Figure~\ref{Fig_06}  shows the covariance of the pair 
\textcolor{black}{$(\sin^2\theta_{23},\,|\Delta m^2|)$}. 
In this case, 
there is a marked preference of SBL reactor data for relatively ``high'' values of 
\textcolor{black}{$|\Delta m^2|$} 
($\pm 2\sigma$ error bars), 
as compared with LBL accelerator data. A compromise is more easily reached for NO, featuring a smaller 
difference between the $\Delta m^2$ values derived from SBL reactor and LBL accelerator data. 
This explains why the preferred mass ordering flips from inverted to normal, when 
passing from Fig.~\ref{Fig_01} to Fig.~\ref{Fig_02}; see also \cite{deSalas:2020pgw,Esteban:2020cvm}.
For the same reason, maximal values of $\theta_{23}$ (corresponding to the lowest values of 
\textcolor{black}{$|\Delta m^2|$}
allowed by LBL accelerator data) are slightly more disfavored by adding SBL reactor data.
The overall preferences for NO and for nonmaximal $\theta_{23}$ are confirmed by atmospheric data that,
however, move the best fit in NO from the upper to the lower octant. 
We emphasize that SBL reactor data, despite having no direct sensitivity to sign($\Delta m^2$) and $\theta_{23}$,
contribute to constrain (via covariances) these two  variables,  in combination with other datasets.

Figure~\ref{Fig_07} shows the covariance of the pair $(\sin^2\theta_{23},\,\delta)$. The octant
ambiguity leads to two quasi-degenerate best fits, surrounded by allowed regions that merge at $2\sigma$ or $3\sigma$. 
In IO there is rather stable preference for the CP-violating case $\delta \simeq 3\pi/2$ in all data combinations, 
with no significant correlation
with $\theta_{23}$. In NO the allowed $\delta$ range is always larger, and includes the
CP-conserving case $\delta \simeq \pi$ at $2\sigma$; moreover, a slight negative correlation between $\delta$ and $\sin^2\theta_{23}$ emerges when adding SBL
reactor data. It is difficult to trace the origin of these null or small covariances, since
the interplay between $\delta$ and $\sin^2\theta_{23}$ (and with $\sin^2\theta_{13}$) is rather subtle, see e.g.\ 
\cite{Minakata:2013eoa,Coloma:2014kca}. 
In any case, the negative correlation emerging in NO slightly amplifies the effect of adding 
atmospheric neutrino data, that prefer
both $\delta\sim 3\pi/2$ and the lower octant of $\theta_{23}$, thus disfavoring  $\delta \simeq \pi$ in a synergic way.
Quantitatively, we find that the CP-conserving value $\delta=\pi$ is disfavored 
at 90\% C.L.\ (or $\sim\!1.6\sigma$, see Fig.~\ref{Fig_03}), while recent  analyses  not
including SK-IV atmospheric data allowed this value at $<1\sigma$ \cite{deSalas:2020pgw,Esteban:2020cvm}.
Although these covariance effects are admittedly small in current data, they are expected
to grow with increasing statistics and accuracy
in LBL accelerator experiments, whose results we comment in more detail through the so-called  bi-event plots, 
derived from  bi-probability plots.

\newpage

\begin{figure}[t]
\begin{minipage}[c]{0.85\textwidth}
\includegraphics[width=0.83\textwidth]{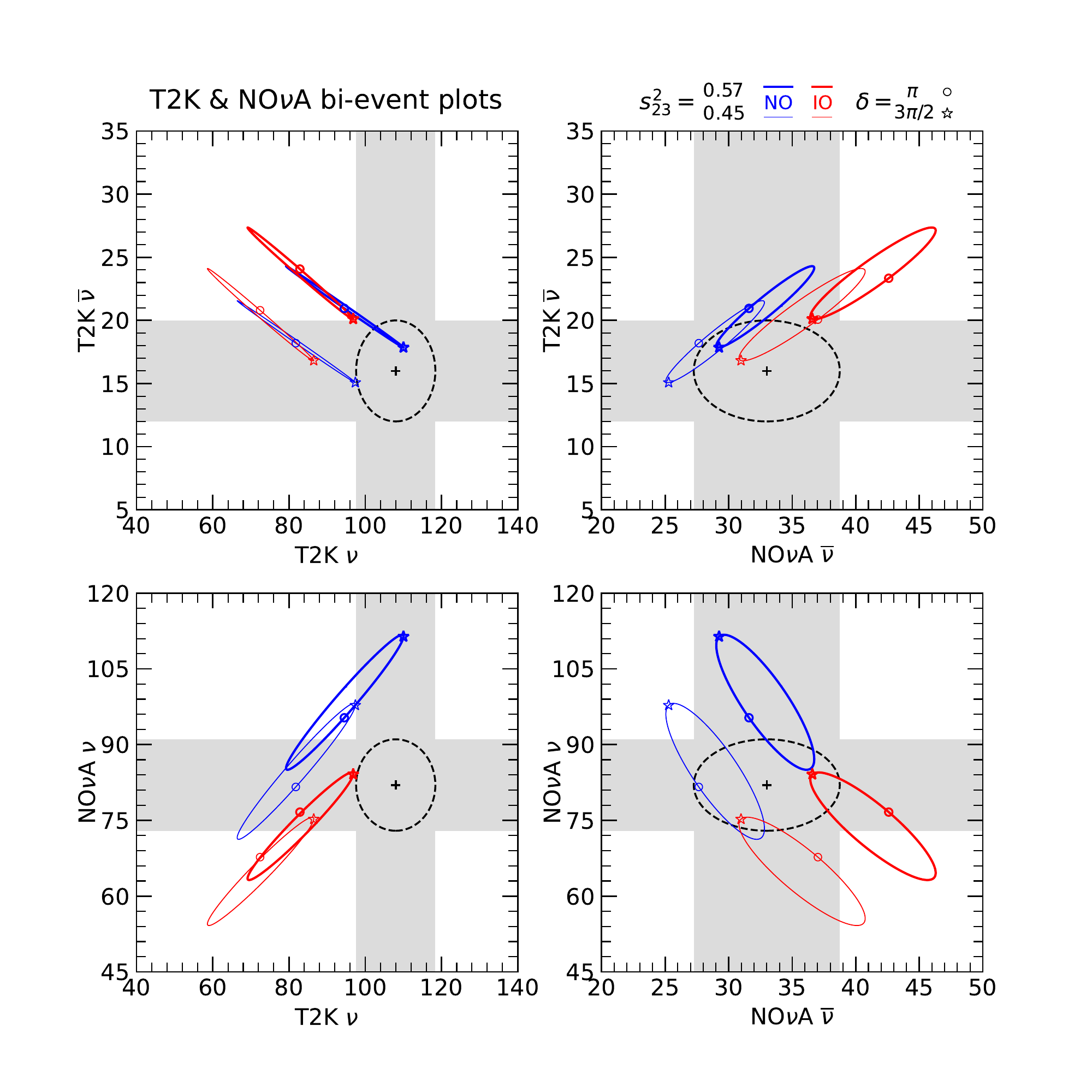}
\vspace*{-5.mm}
\caption{\label{Fig_08}
\footnotesize 
Bi-event plots: Total number of $\nu$ and $\overline\nu$ appearance events 
for T2K and NOvA, in four possible combinations. The slanted ellipses represent the theoretical expectations for NO (blue) and IO (red),
and for two representative values of $\sin^2\theta_{23}$: 0.45 (lower octant, thin ellipses) and 0.57 (upper octant, thick ellipses). The CP-conserving value $\delta=\pi$ and the CP-violating value $\delta=3\pi/2$
are marked as a circle and a star, respectively. Each gray band represents one datum with its $\pm1\sigma$ statistical error (from \cite{T2K2020,NOvA2020}); the combination of any two data provides a (black dashed) $1\sigma$ error ellipse, whose center is marked by a cross.  See the text for details.}
\end{minipage}
\vspace*{-2mm}
\end{figure}

Bi-probability plots, charted by the $\nu_\mu\to\nu_e$ and $\overline\nu_\mu\to\overline\nu_e$ appearance probabilities in LBL accelerator experiments at fixed neutrino energy,
display the cyclic dependence on $\delta$ through ellipses 
\cite{Minakata:2001qm} and help to understand parameter degeneracies \cite{Minakata:2002qi}.
See \cite{Kelly:2020fkv} for a related discussion, in the context of recent T2K and NOvA data.  
After integration over energy (weighted by $\nu$ fluxes and cross sections), the probabilities
can be converted into total number of appearance events and thus into  bi-event plots, preserving elliptic shapes
 \cite{Mena:2006uw}. Such theoretical ellipses can be directly compared with the measured number of events; 
see, e.g., the presentations at {\em Neutrino 2020\/} \cite{Nu2020} by T2K \cite{T2K2020} and NOvA \cite{NOvA2020}.  
Although we use the full energy spectra (and not their integrals) in our LBL accelerator data analysis, we
think that bi-event plots can help to highlight some issues emerging in the comparison of current T2K and NOvA data.

Figure~\ref{Fig_08} show the $\nu$ and $\overline\nu$ appearance events 
for T2K and NOvA, in four possible combinations. The grey bands and the black ellipses represent the data with their 
$1\sigma$ statistical errors, while the colored ellipses represent the theoretical expectations for NO (blue) and IO (red),
for two representative values of $\theta_{23}$ in the lower octant (thin) or upper octant (thick). Two representative values
of $\delta$ ($\pi$ and $3\pi/2$) are also marked on each ellipse. 

We first consider the two experiments separately, as shown in 
the upper left panel for T2K ($\nu$ vs $\overline\nu$) and in the lower right panel for NOvA ($\nu$ vs $\overline\nu$). In T2K, the best agreement of theory and data is reached for NO; in this ordering, there is a clear preference for $\delta=3\pi/2$, and a slight preference for the upper
octant of $\theta_{23}$. In NOvA, all the four
theoretical ellipses are close to the experimental one, but the overlap is larger in NO; in this ordering,
there is a preference for $\delta=\pi/2$ with respect to $3\pi/2$, with no significant distinction of the $\theta_{23}$ octants. 

We then rearrange exactly the same information (both data and predictions)
by combining T2K and NOvA separately in the $\nu$ and $\overline\nu$ channels,
as shown in the upper right panel for $\overline\nu$, and in the lower left panel for $\nu$. 
In both plots, the best agreement of data and theory is now reached for IO. In such ordering,
there is  a clear preference for $\delta=3\pi/2$, as well as for the upper (lower) octant in the $\nu$ ($\overline\nu$) channel.   

In conclusion, Fig.~\ref{Fig_08} shows that, as far as the three oscillation unknowns are concerned 
(mass ordering, $\theta_{23}$ octant, CP symmetry), separate and combined T2K and NOvA data provide us with different indications, signalling a ``tension'' between such results; see also the
discussion in \cite{deSalas:2020pgw,Esteban:2020cvm,Kelly:2020fkv}.
\textcolor{black}{Ultimately, the tension reflects the fact that NOvA (T2K) observes relatively symmetric (asymmetric) rates of $\nu$ and 
$\overline\nu$ in their current appearance data.}

\subsection{Remarks}
\label{sec:remarks}

The current hints about the 
three oscillation unknowns are less converging and more fragile than in the recent past 
\cite{Capozzi:2020}, due to the T2K-NOvA tension. As for its origin,
Fig.~\ref{Fig_08} shows that possible statistical fluctuations of the data (at the level of one or two standard deviations) might play a role. However, it makes sense to  
speculate if there is more than just statistics behind the tension.  
One possibility is to invoke nonstandard neutrino interactions, that would induce
different effects along the T2K and NOvA baselines \cite{Chatterjee:2020kkm,Denton:2020uda}. 
Barring new physics beyond the $3\nu$ paradigm, we surmise that standard
interactions of neutrinos in nuclei might also play a systematic role.
   
It is widely recognized that both the total and the differential neutrino cross sections in nuclei are not known 
accurately enough for the purposes of LBL accelerator experiments \cite{Mosel:2016cwa,Alvarez-Ruso:2017oui}. Roughly speaking,
normalization and energy reconstruction uncertainties in $\nu_\mu$ cross sections 
affect the measurement of $\theta_{23}$ and $\Delta m^2$, respectively, while
systematics on the relative $\nu_\mu/\nu_e$ and $\nu/\overline\nu$ cross sections  affect the LBL experimental sensitivity
to $\theta_{13}$, $\delta$ and the mass ordering. 
Note, e.g., that a 1\% systematic error on the reconstructed neutrino energy $E^\mathrm{rec}_\nu$ is transferred to $\Delta m^2$ 
via the leading $\Delta m^2/E$ dependence of the oscillation phase. The formal $\sim\!1 \%$ error on $\Delta m^2$ 
emerging from the global fit (Table~\ref{Tab:Synopsis}) implicitly posits 
that energy reconstruction errors are known at sub-percent level and are independent in different experiments,
which may be an optimistic representation of the current uncertainties.  

With increasingly higher statistics and accuracy, cross-section
systematics may start to affect both known and unknown oscillation parameters
extracted from detailed energy spectra.  Possible parameter biases have been shown to arise by swapping 
different cross section models in simulations of prospective LBL data   
\cite{Coloma:2013rqa,Coloma:2013tba,Benhar:2015wva}.  We remark that
a percent-level bias on $\Delta m^2$ as measured at LBL
accelerators, in comparison with the $\Delta m^2$ measurement at reactors, might alter the current combined preference for NO
(see Fig.~\ref{Fig_06} and related discussion).
Although all these effects can be reduced 
by tuning interactions models to cross-section data from near detectors \cite{T2K2020,NOvA2020}, 
as a matter of fact T2K and NOvA use two different such models, while no single  model or neutrino generator
can be currently tuned to agree with world cross-section data \cite{Alvarez-Ruso:2017oui,Barrow:2020gzb}.

Summarizing, the global analysis of current data shows a subtle interplay between the known oscillation parameters 
\textcolor{black}{$(|\Delta m^2|,\,\sin^2\theta_{23})$} 
and the three unknowns $(\delta,\,\mathrm{sign}(\theta_{23}-\pi/4),\,\mathrm{sign}(\Delta m^2))$, 
as discussed through Figs.~\ref{Fig_05}--\ref{Fig_07}. Although there are overall hints in favor of NO
(at $2.5\sigma$), CP violation (at $1.6\sigma$) and lower $\theta_{23}$ octant  (at $1.6\sigma$), the  T2K--NOvA tension  (Fig.~\ref{Fig_08}) warrants some caution. Neutrino interaction systematics might affect all these parameters, in a way that escapes control in
global fits by external users, since the complexity of the near-to-far analysis chain
can be handled only by the experimental collaborations. 
It is thus encouraging that T2K and NOvA are planning a joint analysis \cite{JOINT}. In this context, we practically
suggest that these two experiments try to swap interaction models or neutrino generators in their separate simulations, 
so as to gauge the relative size of cross-section systematics and tuning effects, before attempting a combined fit. 
In general, we suggest that experiments sharing potential relevant systematics 
(e.g., neutrino fluxes $\Phi_\alpha$ and interactions in water $\sigma_\beta$ for atmospheric neutrinos) collaborate on 
a detailed comparison of such uncertainties and possibly towards joint data analyses. 
Of course, experimental developments on $\Phi_\alpha$ and $\sigma_\beta$ should be accompanied by 
corresponding advances in nuclear theory.

\section{Nonoscillation data, analysis methods and results}
\label{Sec:Nonosc}

In this section we introduce recent nonoscillation data that were not included in our previous work \cite{Capozzi:2020}, together with the methodology used for their analysis in terms of the three absolute mass observables $(m_\beta,\,m_{\beta\beta},\Sigma)$. We include the latest $m_\beta$ upper bounds  from the KATRIN experiment \cite{KATRIN2021} and 
introduce a method to combine the latest $0\nu\beta\beta$ constraints in terms 
of upper bounds on $m_{\beta\beta}$, including correlated uncertainties on their nuclear matrix elements. We also enlarge the ensemble of cosmological datasets presented in \cite{Capozzi:2020} in the light of the current lively discussion
on tensions in cosmology \cite{DiValentino:2021izs,DiValentino:Snowmass,Challenge2021}, that might suggest possible inconsistencies among different data (or alterations of the standard $\Lambda$CDM model).  In particular, we consider an ``alternative'' dataset, that is exempt from the Planck lensing anomaly, at the price of being restricted to recent cosmic microwave background (CMB) anisotropy observations from ACTPol-DR4  \cite{Aiola:2020azj} plus WMAP9 \cite{Bennett:2012zja} 
and selected Planck data \cite{Aghanim:2018eyx,Aghanim:2019ame,Aghanim:2018oex}. The alternative option provides a nonzero best fit and
more conservative upper bounds on
$\Sigma$, as compared with the ``default''  
dataset described in \cite{Capozzi:2020}. 
We shall highlight the different sensitivities to $\Sigma$ and to the mass ordering in the default and alternative options, as examples of admissible cosmological variants with rather different impact on global neutrino data analyses.

\subsection{Single beta decay and constraints on $m_\beta$}

The KATRIN $\beta$-decay experiment has recently released the results of the second campaign  of measurements \cite{KATRIN2021}. In combination with the results of the  first campaign \cite{Aker:2019uuj}, they constrain
at $1\sigma$ the effective  squared mass $m^2_\beta$  as:
\begin{equation}
\label{eq:KATRIN}
m_\beta^2 = 0.1 \pm 0.3\ \mathrm{eV}^2,
\end{equation}
with an approximately gaussian distribution  around the best fit, currently in the physical region
$m^2_{\beta}>0$ \cite{KATRIN2021} (
while it was negative in the first campaign \cite{Aker:2019uuj}). The upper bound at 90\% C.L.\ ($\sim\! 1.6\sigma$) 
corresponds to $m^2_\beta<0.6$~eV$^2$ or $m_\beta<0.8$~eV, representing the first  constraint on the effective
$\beta$-decay mass in the sub-eV range \cite{KATRIN2021}. Note that variants of the statistical analysis may lead to small 
differences in the second significant digit of $m^2_\beta$ or $m_\beta$ \cite{KATRIN2021},
not considered herein.  We implement the datum in Eq.~(\ref{eq:KATRIN}) via a contribution $((x-0.1)/0.3)^2\in \chi^2$, where $x=(m_\beta/\mathrm{eV})^2$. 

\subsection{Neutrinoless double beta decay and constraints on half-lives and $m_{\beta\beta}$}

Neutrinoless double beta decay \cite{PDG6} can be considered as the process of creation of two matter particles (electrons) \cite{Vissani:2021gdw}, occurring if neutrinos are of Majorana type \cite{Bilenky:2019gzn}. 
Within the $3\nu$ paradigm, the decay half-life $T$ is given by
\begin{equation}
\label{eq:2beta}
\frac{1}{T_i} = G_i |M_i|^2 m^2_{\beta\beta} = S_i
\end{equation}
where the index $i$ labels the $0\nu\beta\beta$ nuclide, characterized by a phase space $G_i$ and a nuclear matrix element (NME) $M_i$, while $m_{\beta\beta}$ is the effective Majorana mass.  The inverse half life $S_i=1/T_i$ represents, up to a constant factor, the observable decay rate or signal strength.

Current experiments are consistent with null signal ($S=0$), placing lower limits on $T$ and upper limits on $m_{\beta\beta}$ \cite{Formaggio:2021nfz} via Eq.~(\ref{eq:2beta}). In deriving separate and combined limits on $m_{\beta\beta}$, two issues arise: (1) experimental results are often given in terms of 
90\% C.L.\ bounds on $T$ (say, $T>T_{90}$), with little or no information on the probability distribution of $T$ (or of $S$);
(2) theoretical uncertainties on the NME are rather large (and correlated among nuclides, e.g., via the axial coupling)
\cite{Engel:2016xgb}. See e.g.\ 
\cite{Biller:2021bqx} for a recent discussion. 
We describe below our approach to these issues, in order to build first a probability distribution for half-lives $T_i$ and then, including NME's, for the effective mass $m_{\beta\beta}$.

We limit our analysis to experiments placing limits $T_{90}> 10^{25}$~y, namely: 
GERDA \cite{Agostini:2020xta}
and MAJORANA \cite{Alvis:2019sil}
for $^{76}$Ge;
CUORE \cite{Adams:2021rbc}
for $^{130}$Te;
KamLAND-Zen 400 \cite{KamLAND-Zen:2016pfg}
and 800 (preliminary) \cite{Gando:2020cxo} 
and EXO-200 \cite{Anton:2019wmi}
for $^{136}$Xe. For each experiment we need not only a limit ($T_{90}$) but the  probability distribution of $T_i$ or, equivalently, a function $\Delta\chi^2(S_i)$. 
Unfortunately, such detailed information is not provided by current
experiments in an explicit or user-friendly way; see \cite{Biller:2021bqx,Caldwell:2017mqu} for recent attempts to 
parametric reconstructions.

We adopt a $\Delta\chi^2(S_i)$ parametrization inspired by \cite{Caldwell:2017mqu} and based on the following considerations.
In $0\nu\beta\beta$ searches with zero background and nearly null results, the likelihood $\cal L$ of a signal $S>0$ should be a poissonian with a scaling coefficient $\mu$ ($\cal L \sim \mathrm{exp}(-\mu S)$), leading to a linear
dependence on $S$ ($\Delta \chi^2 \sim \ln {\cal{L}} \propto S$) \cite{Biller:2021bqx}. In $0\nu\beta\beta$ searches with nonnegligible background subtraction, the dependence is expected to be nearly  gaussian ($\Delta \chi^2 \propto (S-S_0)^2$),
where the best-fit signal $S_0$ may fall either in the physical region $(S_0\geq 0)$
or in the unphysical one $(S_0<0)$. All these limiting cases can be covered by a quadratic form
\begin{equation}
\label{eq:abc}
\Delta\chi^2(S_i) =  a_i\, S_i^2 + b_i\,S_i + c_i  \ ,
\end{equation}
as previously advocated  in \cite{Caldwell:2017mqu} on
an empirical basis. Note that the offset $c_i$ is set by the condition that $\Delta\chi^2\geq 0$ 
in the physical region $S_i\geq0$. In particular, for $a_i>0$, a $\Delta\chi^2$ minimum in the physical region implies
$b_i\leq 0$ and $c_i=b_i^2/4a_i$, while in the unphysical one it implies $b_i> 0$ and $c_i=0$. For $a_i=0$ one recovers the linear  limit, that implies $b_i>0$ and $c_i=0$. The case $a_i<0$ is never realized.

In order to assess the coefficients $(a_i,\,b_i,\,c_i)$, we have carefully sifted 
the information contained in the experimental publications 
\cite{Agostini:2020xta,
Alvis:2019sil,
Adams:2021rbc,
KamLAND-Zen:2016pfg,
Gando:2020cxo,
Anton:2019wmi} and in
available PhD theses conducted within EXO-200
\cite{Jewell,Ziegler}, KamLAND-Zen 400 \cite{Sayuri} and KamLAND-Zen 800 (preliminary) \cite{Ozaki}.
We find that the linear approximation  advocated in \cite{Biller:2021bqx} for GERDA, can be 
roughly applied also to MAJORANA (up to subleading corrections at small $S_i$, 
neglected herein). The other experimental bounds on $T_i$ can be reasonably
approximated by parabolic $\Delta\chi^2$ curves, setting the various coefficients $(a_i,\,b_i,\,c_i)$. We also
require that our 90\% C.L.\ limit $\Delta \chi^2(S_{90})=2.706$ reproduces the $T_{90}$ limit reported by each 
experiment, up to their quoted significant digits. 
We have further checked that, by shifting the minimum of each $\Delta\chi^2$ function from $S=S_0$ to $S=0$,
the corresponding 90\% C.L.\ limits are in reasonable agreement with the reported sensitivities
for the null hypothesis.
We are thus confident that current experimental results are fairly well represented by our $\Delta\chi^2$'s.
Finally, we combine different experiments probing the same nuclide,
by adding up their $\Delta\chi^2$ functions.


\begin{table}[t!]
\centering
\resizebox{.8\textwidth}{!}{\begin{minipage}{\textwidth}
\caption{\label{tab:abc} 
Neutrinoless double beta decay: Details of the adopted parametrization $\Delta\chi^2(S_i)=a_i\, S_i^2 + b_i\,S_i + c_i  
$ for the signal strength $S_i=1/T_i$, expressed
in units of 10$^{-26}$~y$^{-1}$. The first two columns report the nuclide and the
name of the experiment(s). The next three columns report our evaluation of the 
coefficients $(a_i,\,b_i,\,c_i)$, for the various experiments, taken either separately (upper six rows) or in combination
for the same nuclide (lower three rows).  The sixth column reports our 90\% C.L.\ 
($\Delta\chi^2=2.706$) half-life limits $T_{90}$ in units of $10^{26}$~y, to be compared with
the experimentally quoted ones in the seventh column (in the same units). Pertinent references are listed in the last column. 
}
\begin{ruledtabular}
\begin{tabular}{rlrrrccc}
Nuclide 	& Experiment(s) 			& $a_i~~~$ 	& $b_i~~~$ 	& $c_i~~~$ 	& $T_{90}/10^{26}\,\mathrm{y}$ & $T_{90}$ (expt.)
										& References \\ 
\hline
$^{76}$Ge	& GERDA						& 0.000 	& \textcolor{black}{4.867} 	& 0.000 	& 1.800	& 1.8
										& \cite{Agostini:2020xta}\\
$^{76}$Ge	& MAJORANA					& 0.000		& \textcolor{black}{0.731} 	& 0.000		& 0.270 & 0.27 
										& \cite{Alvis:2019sil}\\
$^{130}$Te	& CUORE						& 0.245		& $-0.637$ 	& 0.414		& 0.216 & 0.22 
										& \cite{Adams:2021rbc}\\
$^{136}$Xe	& KamLAND-Zen 400			& 0.540		& 2.374		& 0.000		& 1.065 & 1.07 
										&\cite{KamLAND-Zen:2016pfg,Sayuri} \\
$^{136}$Xe	& KamLAND-Zen 800 prelim.	& 1.006		& $-0.169$	& 0.007 	& 0.580 & 0.58 
										&\cite{Gando:2020cxo,Ozaki}\\
$^{136}$Xe	& EXO-200					& 0.440		& $-0.338$ 	& 0.065		& 0.350 & 0.35 
										& \cite{Anton:2019wmi,Jewell,Ziegler}\\
\hline
$^{76}$Ge	& GERDA + MAJORANA			& 0.000		& 5.598		& 0.000		& 2.070 & \textemdash 
										& This work \\
$^{130}$Te	& CUORE	(same as above)		& 0.245		& $-0.637$ 	& 0.414		& 0.216 &  0.22 
										& \cite{Adams:2021rbc} \\
$^{136}$Xe	& KamLAND-Zen (400 + 800 prelim.) + EXO-200			
										& 1.986		& 1.867		& 0.000		& 1.267 & \textemdash 
										& This work
\\
\end{tabular}
\end{ruledtabular}
\end{minipage}}
\end{table}

Table~\ref{tab:abc} reports our numerical results for the coefficients $(a_i,\,b_i,\,c_i)$ used in Eq.~(\ref{eq:abc}),
for both separate and combined bounds. [We formally keep up to four significant digits, 
to avoid accumulation of round-off errors in the analysis.] By combining GERDA and MAJORANA, 
we evaluate a 90\% C.L.\ limit as high as $T>2.07\times10^{26}$~y for the $^{76}$Ge half life, about 15\% higher than from GERDA alone. A similar improvement is obtained for $^{136}$Xe, reaching $T>1.27\times10^{26}$~y in the
combination of KamLAND-Zen and EXO data. For $^{130}$Te, CUORE alone sets the bound $T>0.22\times10^{26}$~y.

\begin{figure}[b!]
\begin{minipage}[c]{0.85\textwidth}
\includegraphics[width=0.88\textwidth]{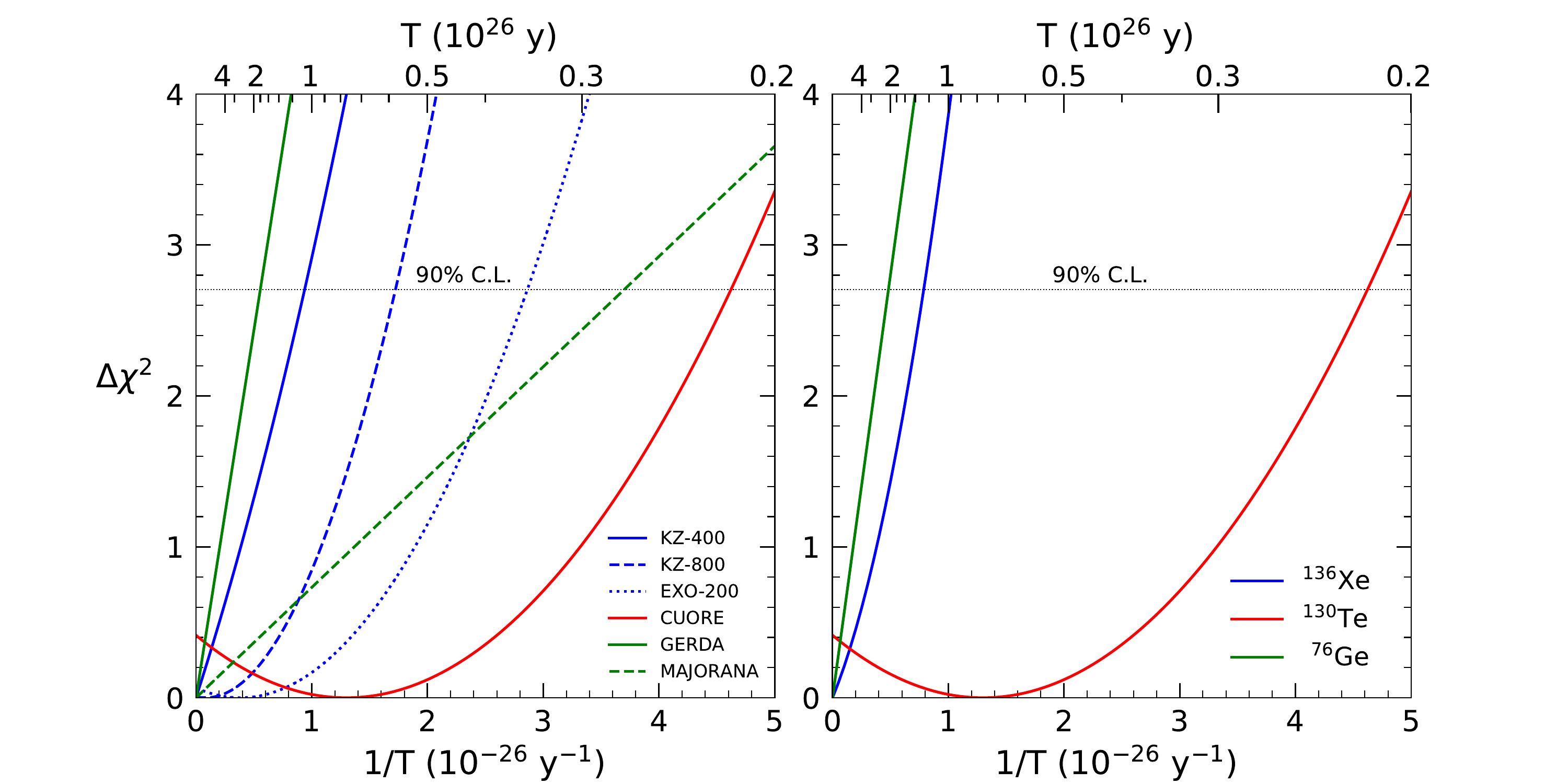}
\caption{\label{Fig_09}
\footnotesize 
Neutrinoless double beta decay: $\Delta\chi^2$ functions as defined in Eq.~(\ref{eq:abc}),
in terms of the half life $T$ (top abscissa) and of $S=1/T$ (bottom abscissa). 
Left and right panels: separate experiments and their combinations for the same nuclide, respectively.
Dotted horizontal lines intersect the curves at 90\% C.L.}
\end{minipage}
\end{figure}

Figure~\ref{Fig_09} shows our parametrized $\Delta\chi^2$ functions
in terms of $1/T=S$ (bottom abscissa) and of $T$ (top abscissa). The left and right panels refer to separate experiments and to combinations for the same nuclide, respectively. The dotted horizontal lines intersect all curves at the 90\% C.L.\ limit
$T_{90}$. Note that the hierarchy of bounds at $90\%$ C.L.\ is not necessarily preserved at different statistical levels, since
some curves cross each other. This is another reason to suggest that the experimental collaborations 
explicitly provide their probability
profiles for $T$, rather than focusing on a single C.L.\ limit ($T_{90}$), that 
provides a poor summary of the data and a limited comparison of different results.

In order to translate constraints from $T_i$ to $m_{\beta\beta}$, one needs to know the phase space factors $G_i$ and
the nuclear matrix elements $|M_i|$ in Eq.~(\ref{eq:2beta}). The $G_i$ can be accurately computed, see
\cite{Deppisch:2020ztt,Mirea:2015nsl} for recent calculations.  The $|M_i|$ embed complex nuclear physics, currently treated
in a variety of approaches that, unfortunately, still carry significant uncertainties despite the 
theoretical progress, see e.g.\ the recent reviews in  \cite{Engel:2016xgb,Coraggio:2020iht,Cirigliano:2020yhp,Dolinski:2019nrj,Ejiri:2019ezh,Vergados:2016hso,DellOro:2016tmg,Barea:2015kwa}. It is 
common practice to select a set of published values $\{M_i\}$ in order to obtain a set of upper  
bounds $\{m_{\beta\beta}\}$, whose 
spread may be taken as indicative of the theoretical uncertainties; see \cite{Biller:2021bqx} for a very recent application.
However, this procedure overlooks significant correlations among the NME uncertainties of different nuclides 
\cite{Faessler:2008xj,Faessler:2013hz,Faessler:2011rv,Lisi:2015yma} 
that, as shown below, are as important as the uncertainties themselves in obtaining conservative bounds.

In a given nuclear model, estimating NME covariances requires massive numerical experiments to generate many $M_i$ variants. To our knowledge, this task has been performed only in \cite{Faessler:2008xj} within the quasiparticle random phase approximation (QRPA), by  varying the axial coupling $g_A$, the short-range correlations, the model basis, and 
the renormalization procedure, while requiring consistency with available $2\nu\beta\beta$ data. Apart 
from the subsequent papers \cite{Faessler:2013hz,Faessler:2011rv,Lisi:2015yma}, and
despite the potential relevance of NME covariance issues  \cite{Engel:2015wha},
we are aware of just one independent
(but preliminary) correlation matrix estimate in a different model \cite{Gautam:2017bgq} 
and of a single statistical analysis \cite{Ge:2017erv} based 
on the correlations in \cite{Faessler:2008xj}.
In the absence of novel estimates of NME covariances, we shall use the only available results of \cite{Faessler:2008xj} at face value. 
We have checked
that, in any case, the NME uncertainties estimated therein are conservative enough to  cover within $\pm2\sigma$  most
(and within $\pm3\sigma$ all)  of the $M_i$ values
compiled in  \cite{Engel:2016xgb,Coraggio:2020iht,Cirigliano:2020yhp,Dolinski:2019nrj,Ejiri:2019ezh,Vergados:2016hso,DellOro:2016tmg,Barea:2015kwa} for the three nuclides.

We summarize and use the results in \cite{Faessler:2008xj} as follows. In order to deal with large $|M_i|$ variations, 
possibly hitting the unphysical range $|M_i|<0$, the (adimensional) $|M_i|$ values are parametrized in terms of the logarithms $\eta_i$,
\begin{equation}
 |M_i| = e^{\eta_i} = e^{\overline \eta_i+\Delta \eta_i} = |\overline M_i|e^{\Delta \eta_i}\ ,  
\end{equation}
where the overlined symbols represent central values, $|\overline M_i|=e^{\overline\eta_i}$. The index $i=1,\,2,\,3$ runs over $^{76}$Ge, $^{130}$Te, $^{136}$Xe.
  The NME variations $\Delta \eta_i$ are endowed with variances $\sigma^2_i$
and a covariance matrix $\sigma_{ij}=\rho_{ij}\sigma_i\sigma_j$, whose inverse defines the weight matrix $w_{ij}=(\sigma_{ij})^{-1}$. For any choice of $m_{\beta\beta}$ and of $\Delta \eta_i$, the signal strength in the
$i$-th nuclide is given by
\begin{equation}
S_i = G_i  |M_i|^2 m^2_{\beta\beta} = q_i m^2_{\beta\beta} e^{2\Delta \eta_i}\ ,  
\end{equation}
where $q_i=G_i |\overline M_i|^2$. The strengths $S_i$ carry two $\Delta \chi^2$ contributions: an experimental one coming from $\Delta \chi^2(S_i)$,   and a theoretical one coming from NME covariances. These contributions are coupled by---and must be minimized over---the three variations $\{\Delta\eta_i\}$
\begin{eqnarray}
\chi^2(m_{\beta\beta}) &=& \min_{\Delta_i}\left(
\sum_{i=1}^3 \Delta\chi^2(S_i)+\sum_{i,j=1}^3 w_{ij}\, \Delta\eta_i\Delta\eta_j\right)\\
&=& \min_{\Delta_i}\left( \sum_{i=1}^3 \left( a_i q^2_i m^4_{\beta\beta}e^{4\Delta\eta_i}
+b_i q_i m^2_{\beta\beta}e^{2\Delta\eta_i} + c_i\right) + \sum_{i,j=1}^3 w_{ij}\, \Delta\eta_i\Delta\eta_j\right)\ ,
\end{eqnarray}
where the first line holds in general, while the second line refers to the specific parametrization in
Eq.~(\ref{eq:abc}). 
Minimization yields three coupled equations in the three $\Delta\eta_i$ unknowns, to be solved numerically. 
Neglecting NME correlations (or errors)
amounts to setting $\rho_{ij} = \delta_{ij}$ (or $\Delta\eta_i=0$). Bounds for a single nuclide are obtained by selecting a specific index $i$. Subtraction of a $\chi^2_{\min}$ offset (if any) yields
the desired $\Delta\chi^2(m_{\beta\beta})$. Table~\ref{tab:Old} reports the values of $\sigma_{ij}$ and $q_i$ used herein.

\begin{figure}[t!]
\begin{minipage}[c]{0.85\textwidth}
\includegraphics[width=0.88\textwidth]{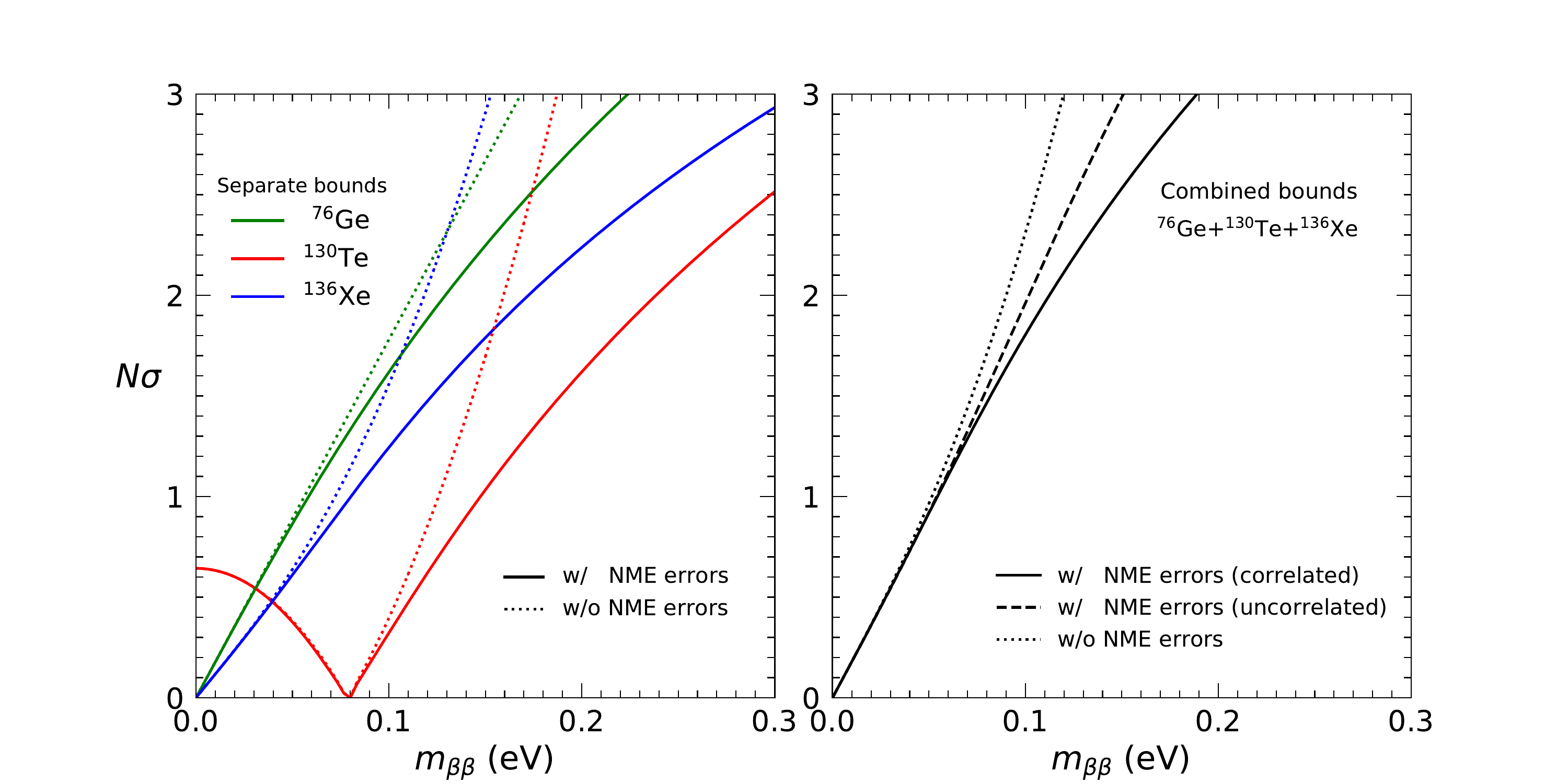}
\caption{\label{Fig_10}
\footnotesize 
Neutrinoless double beta decay: Our estimated bounds on $m_{\beta\beta}$, expressed in terms of $N_\sigma=\sqrt \Delta\chi^2$. 
The left and right panels refer, respectively, to separate and combined bounds from the three nuclides,  
with (solid) or without (dotted) NME uncertainties. In the right panel, 
the case with uncorrelated uncertainties is also shown (dashed).}
\end{minipage}
\end{figure}

Figure~\ref{Fig_10} shows the resulting bounds on $m_{\beta\beta}$ for 
the three nuclides taken separately (left panel) and in combination (right panel). The solid and dotted curves refer to our estimates with and and without NME uncertainties, respectively. Currently, the most constraining results
are obtained by combining $^{76}$Ge data, followed by weaker constraints from $^{136}$Xe and  $^{130}$Te data.
In the
right panel, the case with uncorrelated NME uncertainties is also shown (dashed line).
The effect of correlations is noticeable and leads to more conservative bounds;
in fact, when the NME of different nuclides are positively correlated,
they are more likely to become all smaller at the same time (with respect to the uncorrelated case), allowing larger 
values of $m_{\beta\beta}$. Including correlated errors, we obtain the overall bound
$m_{\beta\beta}<0.11$~eV  at $2\sigma$; the same bound was previously estimated (although in a less refined way) as $m_{\beta\beta}<0.14$~eV in \cite{Capozzi:2020} and as $m_{\beta\beta}<0.18$~eV in \cite{Capozzi:2017ipn}, reflecting the steady experimental progress in the last few years.

\begin{table}[b!]
\centering
\begin{minipage}{.5\textwidth}
\caption{\label{tab:Old} \footnotesize 
Neutrinoless double beta decay: NME covariance matrix $\sigma_{ij}$ 
and auxiliary coefficients $q_i=G_i|\overline M_i|^2$, as derived from the results in
\cite{Faessler:2008xj}; see the text for details. The $q_i$
are given in units of $10^{-26}\,\mathrm{y}^{-1}\, \mathrm{eV}^{-2}$, for $m_{\beta\beta}$ expressed in eV. }
\begin{ruledtabular}
\begin{tabular}{cc|ccc|c}
$i$ & Nuclide  &  & $\sigma_{ij}$ & & $q_i$ \\
\hline
1 & $^{76}$Ge  & 0.0790 &        &        &  56.6 \\
2 & $^{130}$Te & 0.0920 & 0.0135 &        & 210.0 \\
3 & $^{136}$Xe & 0.0975 & 0.1437 & 0.1858 &  73.1 \\
\end{tabular}
\end{ruledtabular}
\end{minipage}
\end{table}

\newpage

We emphasize that the above methodology can be applied
 to $0\nu\beta\beta$ data consistent with either a null signal or a positive detection.
It may include generic  likelihoods  for the half-lives $T_i$ (hopefully provided by the experimental collaborations) 
 and alternative evaluations of the NME covariances (possibly computed in different nuclear models).

\vspace*{-2mm}
\subsection{Cosmology and constraints on $\Sigma$}
\vspace*{-1mm}

In this section we discuss various choices for cosmological data 
combinations, enlarging the set of cases previously considered in \cite{Capozzi:2020}, in order to deal
with known tensions about lensing data. We then focus 
on two specific cases, dubbed as default and alternative options, leading to different implications for $\Sigma$ and 
the sensitivity to mass ordering.  

We remind that in \cite{Capozzi:2020} we
analyzed the following data in various combinations:  the complete Planck 2018 data (Planck) on the
angular CMB temperature power spectrum (TT) 
plus polarization power spectra (TE, EE) \cite{Aghanim:2018eyx,Aghanim:2019ame}
and  lensing reconstruction power spectrum (lensing) \cite{Aghanim:2018oex}; 
a compilation of Baryon Acoustic Oscillation (BAO) measurements, given by data from the 6dFGS~\cite{6dFGS}, SDSS MGS~\cite{mgs}, and BOSS DR12~\cite{bossdr12} surveys; and the Hubble constant datum [$H_0$({\tt R19})] from HST observations of Cepheids in the Large Magellanic Cloud measurements \cite{Riess:2019cxk}. We assume the standard 6-parameter $\Lambda$CDM model augmented with nonzero neutrino masses ($\Lambda$CDM+$\Sigma$) and, in some cases, we add
an extra empirical parameter $A_{\mathrm{lens}}$ to marginalize over the excess amount of gravitational lensing 
emerging in the fits of the Planck data \cite{Aghanim:2018eyx};
\textcolor{black}{see also \cite{RoyChoudhury:2019hls} for a similar approach.}

Here we also consider an alternative way to deal with the Planck lensing problem, in the light of the wider debate
about data tensions in the standard $\Lambda$CDM model \cite{DiValentino:2021izs,Challenge2021}. In particular,
rather than introducing an additional parameter to account for unknown systematics, one may consider
restricting the analysis to alternative CMB datasets not affected by internal or mutual tensions. 
A possibility is offered by the combination of the recent CMB anisotropy data from ACTPol-DR4
 \cite{Aiola:2020azj} with WMAP9 data  \cite{Bennett:2012zja}, together with a 
Planck-derived conservative gaussian prior  on the optical depth to reionization $\tau$
[$\tau_\mathrm{prior}=0.065 \pm 0.015$]. The ACT Collaboration explored this combination in  \cite{Aiola:2020azj},
finding no evidence for a lensing anomaly, and obtaining a relaxed
upper bound on the total neutrino mass, $\Sigma < 1.2$~eV at $2\sigma$. 
Using the same data combination, we get an upper bound in excellent agreement, $\Sigma<1.21$~eV; interestingly, we also find that the best fit is at $\Sigma\simeq 0.70$~eV, with no significant difference between NO and IO. If we replace the gaussian prior on $\tau$
with the actual Planck polarization data at large angular scale (dubbed Planck LowE) 
\cite{Aghanim:2019ame}, that directly constrain the value of $\tau$, we obtain a slightly stronger upper bound  $\Sigma<1.12$~eV (with a best fit at 
$\Sigma\simeq 0.58$~eV). Finally, in order to achieve a more complete combination of CMB data not in tension with each other,
 we further add to ACT and WMAP the independent Planck lensing reconstruction results  
\cite{Aghanim:2018oex}, that do not suffer of the lensing anomaly. In this case
the upper bound becomes  $\Sigma<0.96$~eV, with a best fit at $\Sigma\simeq 0.54$~eV, as obtained with CMB-only data.
Of course, by including additional data, such as BAO measurements, the upper bound on $\Sigma$ would 
be pushed back to $\sim 10^{-1}$~eV (or even below as shown in \cite{DiValentino:2021hoh}), 
at the price of a significant mutual tension among different datasets;
these fits, that would be more comprehensive but also more discordant, are not further discussed herein.

\newpage 

Table~\ref{Tab:Cosmo} reports, for convenience, the bounds on $\Sigma$ for both the cosmological inputs
considered in  \cite{Capozzi:2020} (cases 0--9) and the new ones discussed above (cases 10--12). These
inputs can be roughly divided into two categories: a first group (cases 0--6) where one includes at face value Planck CMB results (plus
other data) despite the Planck lensing anomaly, obtaining
relatively strong upper bounds on $\Sigma$ and a noticeable sensitivity to mass ordering; 
and a second group (cases 7--12), where one ``solves'' the lensing anomaly (by either adding an extra model parameter
or by considering alternative CMB data), with weaker upper bounds on $\Sigma$ and no significant sensitivity
to NO vs IO. Hereafter we focus on just two representative cases for these two different categories, namely, case \#3 (dubbed ``default'' as in \cite{Capozzi:2020}) and case \#12 (dubbed ``alternative'').

\begin{table}[t]
\resizebox{.78\textwidth}{!}{\begin{minipage}{\textwidth}
\caption{\label{Tab:Cosmo} 
\footnotesize  Results of the global $3\nu$ analysis of cosmological data 
within the standard $\Lambda\mathrm{CDM}+\Sigma$ model (possibly augmented with the $A_\mathrm{lens}$ parameter). 
The inputs numbered from 0 to 9 are the same as in \cite{Capozzi:2020},  and refer to various combinations
of the Planck 2018 angular CMB temperature power spectrum (TT) 
plus polarization power spectra (TE, EE), lensing potential power spectrum (lensing),  Barion Acoustic Oscillations (BAO), and the Hubble constant from HST observations of Cepheids in the Large Magellanic Cloud, $H_0$({\tt R19}).
The inputs numbered from 10 to 12 are new and refer to ACTPol-DR4 and WMAP9 data, in combination with  
a prior on optical depth to reionization ($\tau_\mathrm{prior}$),  
Planck polarization data at large angular scale (lowE), and lensing data. 
For each case we report the $2\sigma$ upper bound on the sum of $\nu$ masses $\Sigma$ (marginalized over NO and IO), together
with the $\Delta\chi^2$ difference between IO and NO, using cosmology only. In the last two columns, we report the 
same information as in the previous two columns, but using  cosmological data plus $ m_\beta$ and  $m_{\beta\beta}$ constraints. The specific cases numbered 3 and 12 are dubbed as default
and alternative, see the text for details.}
\vspace*{0mm}
\centering
\begin{ruledtabular}
\begin{tabular}{cllcc|cc}
\multicolumn{3}{l}{Cosmological inputs for nonoscillation data analysis}  & \multicolumn{2}{c|}{Results: Cosmo only} & \multicolumn{2}{c}{Cosmo + $m_\beta$ + $m_{\beta\beta}$}  \\[1mm]
\# & Model & Data set & $\Sigma$ ($2\sigma$)   & $\Delta\chi^2_\mathrm{IO-NO}$ & $\Sigma$ ($2\sigma$)   & $\Delta\chi^2_\mathrm{IO-NO}$ \\[1mm]
\hline
 0 & $\Lambda\mathrm{CDM}+\Sigma$					& Planck {\scriptsize TT,\,TE,\,EE} 									& $<0.34$ eV & $ 0.9$ & $<0.32$ eV & $ 1.0$   \\
\hline
 1 & $\Lambda\mathrm{CDM}+\Sigma$					& Planck {\scriptsize TT,\,TE,\,EE} + lensing							& $<0.30$ eV & $ 0.8$ & $<0.28$ eV & $ 0.9$  \\
 2 & $\Lambda\mathrm{CDM}+\Sigma$ 					& Planck {\scriptsize TT,\,TE,\,EE} + BAO								& $<0.17$ eV & $ 1.6$ & $<0.17$ eV & $ 1.8$  \\
 3 & $\Lambda\mathrm{CDM}+\Sigma$ 					& Planck {\scriptsize TT,\,TE,\,EE} + BAO + lensing						& $<0.15$ eV & $ 2.0$ & $<0.15$ eV & $ 2.2$  \\
\hline
 4 & $\Lambda\mathrm{CDM}+\Sigma$					& Planck {\scriptsize TT,\,TE,\,EE} + lensing + $H_0$({\tt R19})	& $<0.13$ eV & $ 3.9$ & $<0.13$ eV & $ 4.0$  \\
 5 & $\Lambda\mathrm{CDM}+\Sigma$ 					& Planck {\scriptsize TT,\,TE,\,EE} + BAO + $H_0$({\tt R19}) 		& $<0.13$ eV & $ 3.1$ & $<0.13$ eV & $ 3.2$  \\
 6 & $\Lambda\mathrm{CDM}+\Sigma$ 		& Planck {\scriptsize TT,\,TE,\,EE} + BAO + lensing + $H_0$({\tt R19}) \ \	& $<0.12$ eV & $ 3.7$ & $<0.12$ eV & $ 3.8$  \\
\hline
 7 & $\Lambda\mathrm{CDM}+\Sigma+A_\mathrm{lens}$	& Planck {\scriptsize TT,\,TE,\,EE} + lensing							& $<0.77$ eV & $ 0.1$ & $<0.66$ eV & $ 0.1$  \\
 8 & $\Lambda\mathrm{CDM}+\Sigma+A_\mathrm{lens}$	& Planck {\scriptsize TT,\,TE,\,EE} + BAO								& $<0.31$ eV & $ 0.2$ & $<0.30$ eV & $ 0.3$  \\
 9 & $\Lambda\mathrm{CDM}+\Sigma+A_\mathrm{lens}$	& Planck {\scriptsize TT,\,TE,\,EE} + BAO + lensing						& $<0.31$ eV & $ 0.1$ & $<0.30$ eV & $ 0.2$  \\
\hline
 10 & $\Lambda\mathrm{CDM}+\Sigma$	& ACT + WMAP + $\tau_\mathrm{prior}$													& $<1.21$ eV & $-0.1$ & $<1.00$ eV & $ 0.1$  \\
 11 & $\Lambda\mathrm{CDM}+\Sigma$	& ACT + WMAP +  Planck lowE																	& $<1.12$ eV & $-0.1$ & $<0.87$ eV & $ 0.1$  \\
 12 & $\Lambda\mathrm{CDM}+\Sigma$	& ACT + WMAP + Planck lowE + lensing															& $<0.96$ eV & $0.0$  & $<0.85$ eV & $ 0.1$  \\
\end{tabular}
\end{ruledtabular}
\end{minipage}}
\end{table}

\begin{figure}[b!]
\begin{minipage}[c]{0.85\textwidth}
\includegraphics[width=0.48\textwidth]{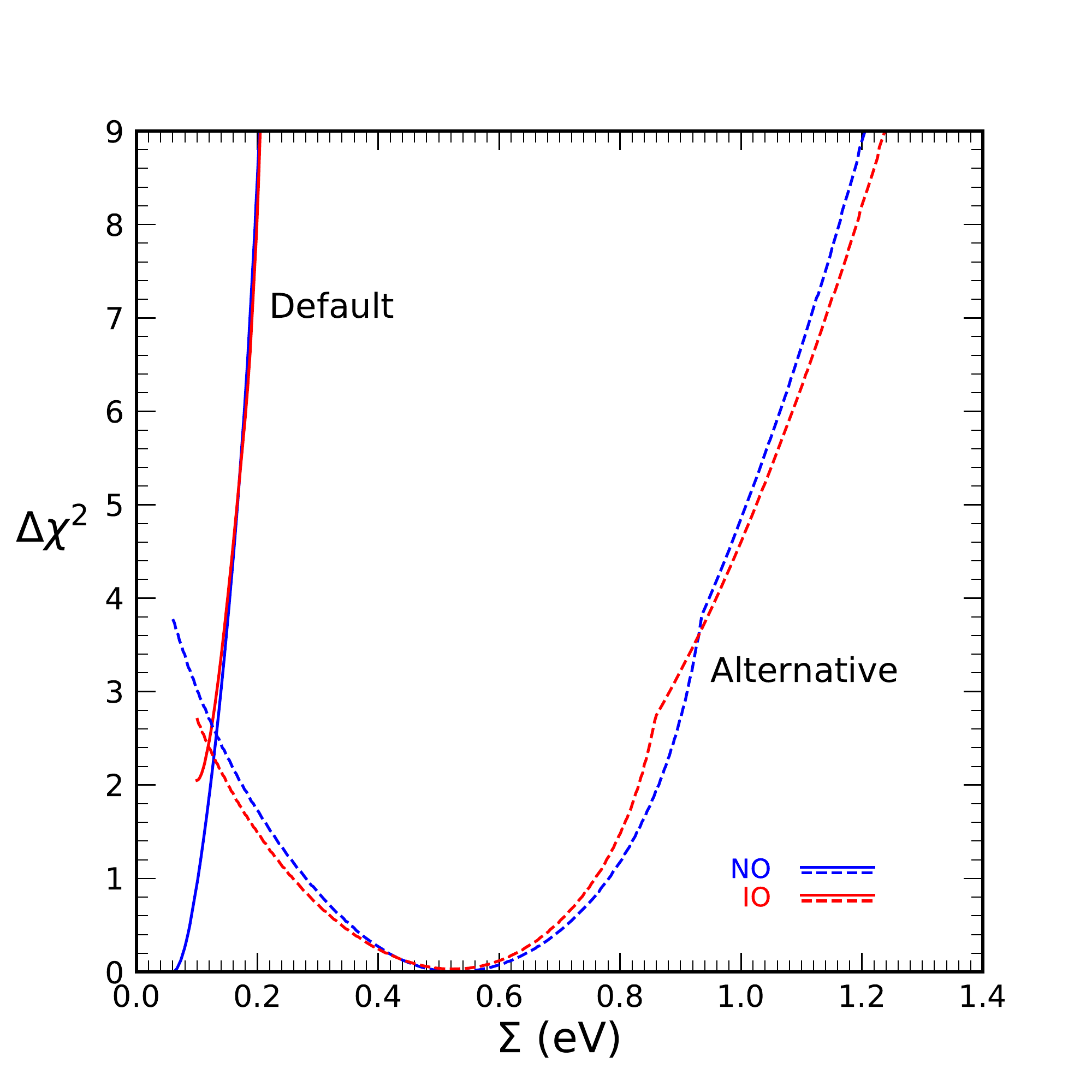}
\vspace*{-3mm}
\caption{\label{Fig_11}
\footnotesize 
Cosmology: $\Delta\chi^2$ curves for NO (blue) and IO (red) for the cases numbered in Table~\ref{Tab:Cosmo} as \#3 (left, solid) and \#12 (right, dashed), taken as representative of default and alternative options, respectively. The cases 
\#10 and \#11 (not shown) would be qualitatively similar to case \#12.}
\end{minipage}
\end{figure}

Figure~\ref{Fig_11} shows the $\Delta\chi^2$ curves for the default and alternative cases. In the default case, cosmological data
push $\Sigma$ towards its lowest physical values in both IO and NO, and favors the latter at the level of $\sim\! 1.5\sigma$. In the alternative case, there
is a  preference $(<2\sigma)$ for higher $\Sigma$ values, with a best fit at $\Sigma\simeq 0.54$~eV and an upper 
limit $\Sigma< 0.96$~eV at $2\sigma$, while there are only minor differences between NO and IO.
Roughly speaking, the alternative case corresponds to a putative cosmological ``signal'' for
neutrino masses, equivalent to $\Sigma \simeq 0.54 \pm 0.22$~eV (with symmetrized $1\sigma$ errors).

\newpage
\textcolor{black}{
We remark that, at fixed $\Sigma$, the small differences between NO and IO fits are due to: a slight sensitivity
of cosmological data to the different ordering of $\nu$ masses at small $\Sigma$; 
the conversion from fit probability densities $P(\Sigma)$ to $\chi^2(\Sigma)$ functions, as the $P$ normalization
covers different physical $\Sigma$ ranges in NO and IO; and, to a lesser extent ($\Delta\chi^2<0.1$), to 
numerical fit inaccuracies for $P\to 0$ (i.e., for high $\chi^2)$. See also the comments in Sec.~{II~C} of \cite{Capozzi:2017ipn}.
} 
   
Summarizing, the two  cases in Fig.~\ref{Fig_11} correspond to two qualitatively different outcomes, 
that might persist in future cosmological data analyses, as opposite examples of the tradeoff between 
completeness and consistency of inputs.
 On the one hand, by 
combining various datasets at face value (regardless of possible tensions), one may obtain strong upper limits,
$\Sigma<O(10^{-1})$~eV, 
with some sensitivity 
to mass ordering (typically in favor of NO, where $\Sigma$ attains its lowest possible values).  
On the other hand, by combining selected (and mutually consistent) data sets, one may end up with
more relaxed upper bounds on $\Sigma$,  possibly shifting the best fit towards $\Sigma\sim (\mathrm{few} \times) 10^{-1}$~eV, at the price of a reduced sensitivity to mass ordering. We think that, at present, both options deserve  to be explored, especially because they imply rather different
outcomes in combination with other (non)oscillation neutrino data. 

\vspace*{-3mm}
\subsection{Results on pairs of nonoscillation observables}
\vspace*{-1mm}

The nonoscillation observables $(m_\beta,\,m_{\beta\beta},\,\Sigma)$ are strongly and positively correlated, 
via their common dependence on the absolute neutrino mass scale. Contrary to the case of oscillation parameters, it is useful to show first the 
results on pairs of observables, and then the projections on single ones.  With respect to our
previous works \cite{Capozzi:2017ipn,Capozzi:2020}, we shall use linear (rather than logarithmic) coordinates 
as, e.g., advocated for the pair $(\Sigma,\,m_{\beta\beta})$ in \cite{DellOro:2016tmg,DellOro:2014ysa}.
Since cosmological data play a major role in constraining the nonoscillation parameter space, we shall discuss the two different options (default and conservative) defined in the previous Section.

\begin{figure}[b!]
\begin{minipage}[c]{0.85\textwidth}
\includegraphics[width=0.45\textwidth]{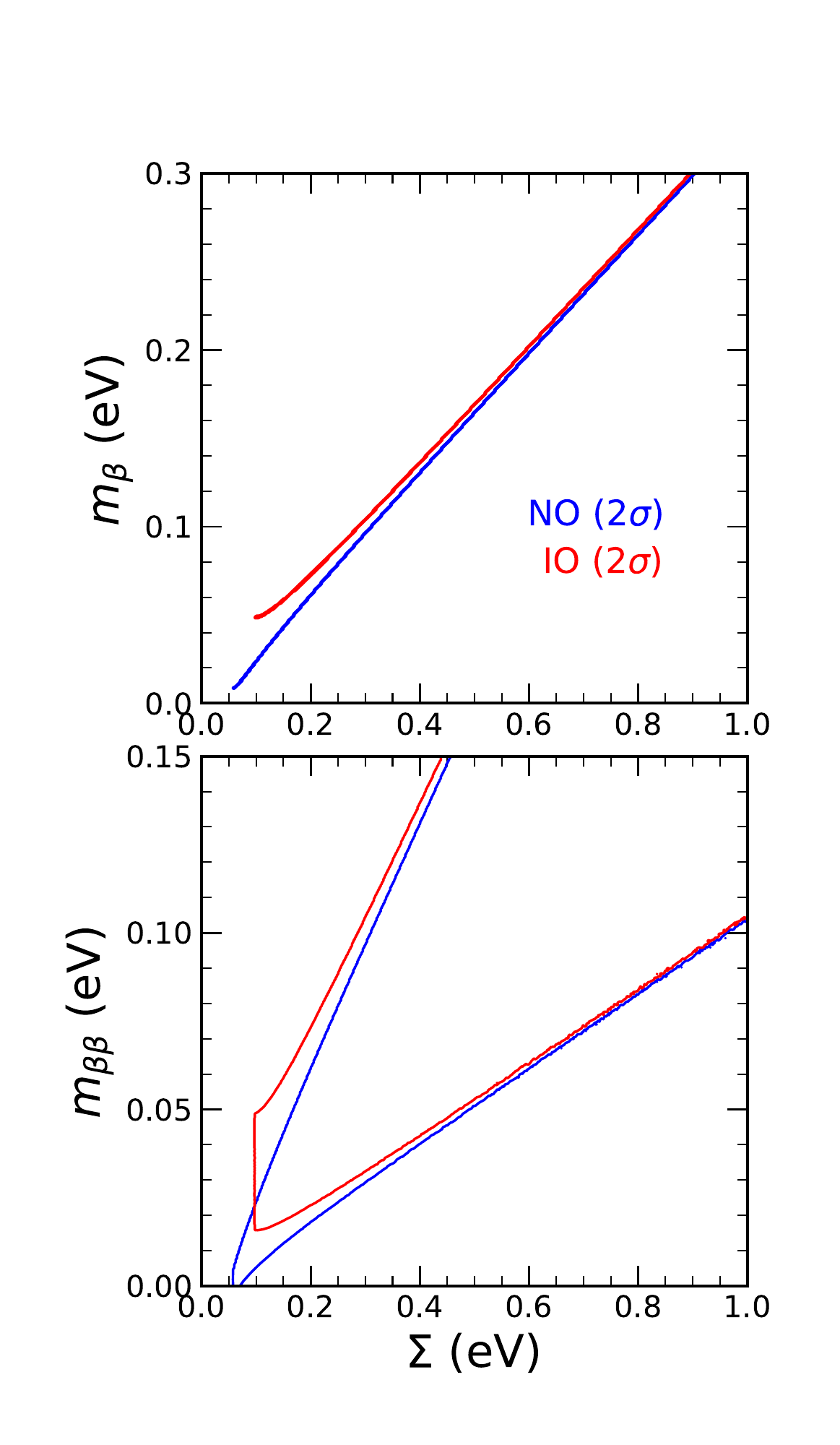}
\vspace*{-4mm}
\caption{\label{Fig_12}
\footnotesize 
Constraints at $2\sigma$ placed by current oscillation data in the planes charted by the nonoscillation observables 
$(\Sigma,\,m_\beta)$ in the top panel, and $(\Sigma,\,m_{\beta\beta})$ in the bottom panel.
The blue and red curves refer to NO and IO, respectively.}
\end{minipage}
\end{figure}

Figure~\ref{Fig_12} shows the correlation bands at $2\sigma$ for the pairs $(\Sigma,\,m_\beta)$ and  $(\Sigma,\,m_{\beta\beta})$ in linear scales, 
including only the constraints from oscillation data, for NO and IO taken separately
(i.e., without the offset $\Delta \chi^2_{\mathrm{{IO}-{NO}}}$). In the top panel, the bands have a  tiny width, reflecting the small fractional errors on the 
oscillation parameters ($\delta m^2,\,\Delta m^2,\,\theta_{12},\,\theta_{13}$) relevant for the pair
$(m_\beta,\,\Sigma)$. In the bottom panel, the widening of the bands is almost entirely due to the unknown Majorana phases
in $m_{\beta\beta}$. See also \cite{Capozzi:2020} and Fig.~2 therein for analogous correlation plots in logarithmic scales.

\vspace*{-5mm}
\newpage

\begin{figure}[t!]
\begin{minipage}[c]{0.85\textwidth}
\includegraphics[width=0.92\textwidth]{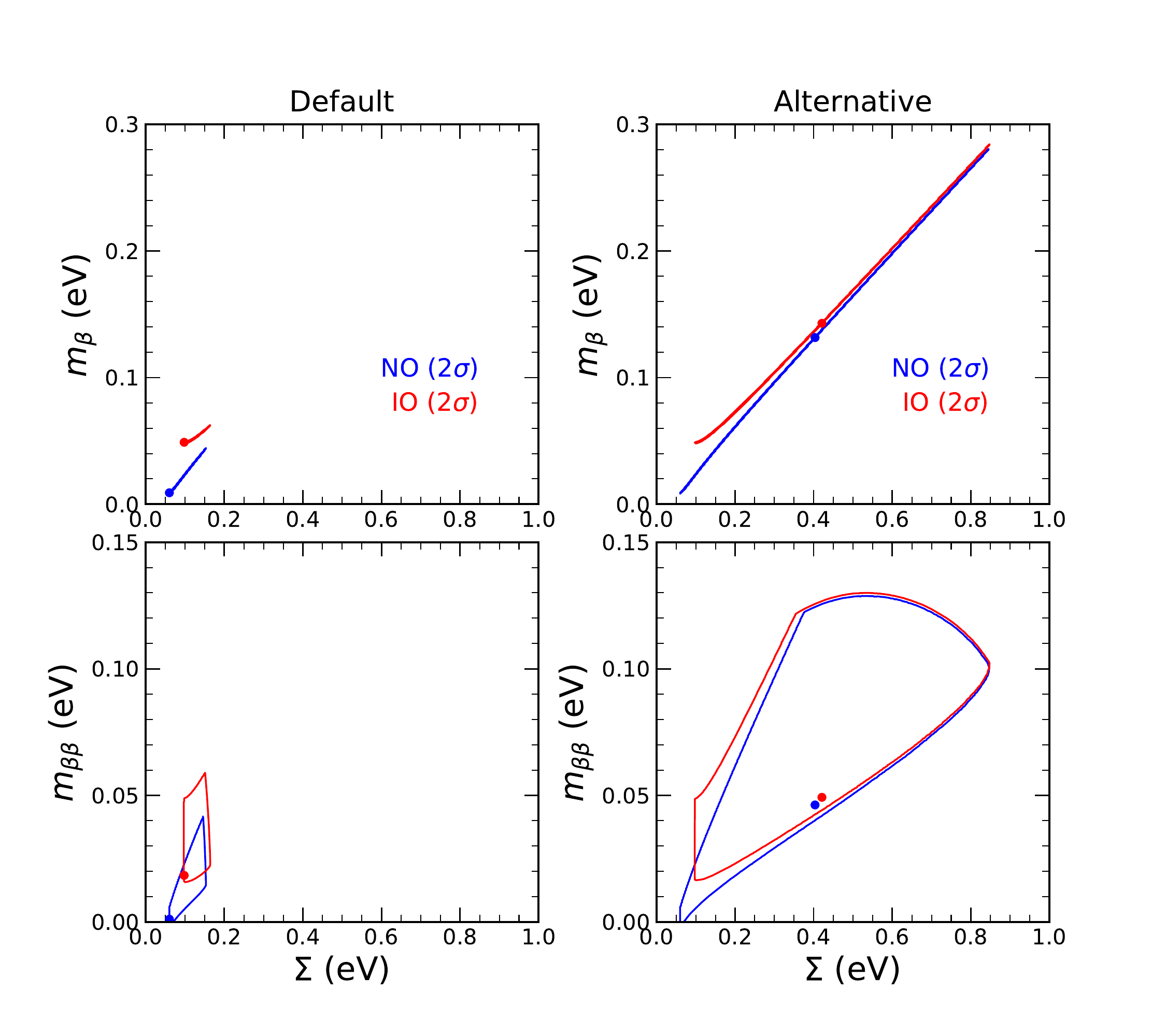}
\caption{\label{Fig_13}
\footnotesize 
Constraints at $2\sigma$ placed by current oscillation data and nonoscillation data 
in the same planes as in Fig.~\ref{Fig_12}, in the default (left panel) and alternative
(right panel) options for cosmological inputs. The dots mark
the best fits. }
\end{minipage}
\end{figure}

\vspace*{10mm}
Figure~\ref{Fig_13} shows how the data from $\beta$ decay, $0\nu\beta\beta$ decay and cosmology further
constrain the $2\sigma$ bands in Fig.~\ref{Fig_12}, for the two cosmological input options considered. In the default case, 
cosmological bounds on $\Sigma$ dominate---via  correlations---the 
constraints on $m_\beta$ and $m_{\beta\beta}$, which are squeezed to the relatively small $2\sigma$
regions around the best fits, located close to the lowest possible values for $\Sigma$ in both NO and IO.
In the alternative case, there is an interplay between cosmological and $0\nu\beta\beta$ data: 
the first would prefer $\Sigma\simeq 0.58$~eV, implying relatively high values 
for the Majorana mass ($m_{\beta\beta}>0.06$~eV, see Fig.~\ref{Fig_12}); however, such values 
are disfavored by $0\nu\beta\beta$ data at  $>1\sigma$ in Fig.~\ref{Fig_10}. A best-fit  
compromise is reached for intermediate values, $\Sigma\sim 0.4$~eV and $m_{\beta\beta}\simeq 0.05$~eV,
surrounded by large $2\sigma$ allowed regions. In the right panel, note that both 
cosmology and  $0\nu\beta\beta$ data constrain the correlations bands from above,
leading to a joint $2\sigma$ bound $\Sigma<0.85$~eV, stronger than the bound from cosmology only  
($\Sigma<0.96$~eV, see also Table~\ref{Tab:Cosmo}). In all cases,
current $\beta$-decay data play a minor role in the overall fit.

The implications of Fig.~\ref{Fig_13} can be summarized  as follows. In the default case, 
it appears that the current KATRIN experiment (probing $m_\beta>0.2$~eV) is
not expected to find any signal, while planned $0\nu\beta\beta$ experiments are expected to probe at
least the region covered by both NO and IO ($m_{\beta\beta}>0.02$~eV). The region covered 
only by NO ($m_{\beta\beta}<0.02$~eV) is more difficult to probe, and becomes eventually   
prohibitive as $m_{\beta\beta}$ vanishes, see e.g. \cite{Dolinski:2019nrj,Pascoli:2007qh,Penedo:2018kpc}.
In the alternative case, a much larger phase space is amenable to $\beta$ decay and $0\nu\beta\beta$ decay
searches. Cosmological searches may find a signal for $\Sigma$ in a wide sub-eV range. 
Neutrinoless double beta decay data might find a signal for $m_{\beta\beta}$ anywhere 
below the current bounds. 
The KATRIN experiment might find a signal in its sensitivity region $(m_\beta>0.2)$~eV,
or at least strengthen significantly the upper bounds on $m_\beta$. 

\newpage

\begin{figure}[t!]
\begin{minipage}[c]{0.85\textwidth}
\includegraphics[width=0.95\textwidth]{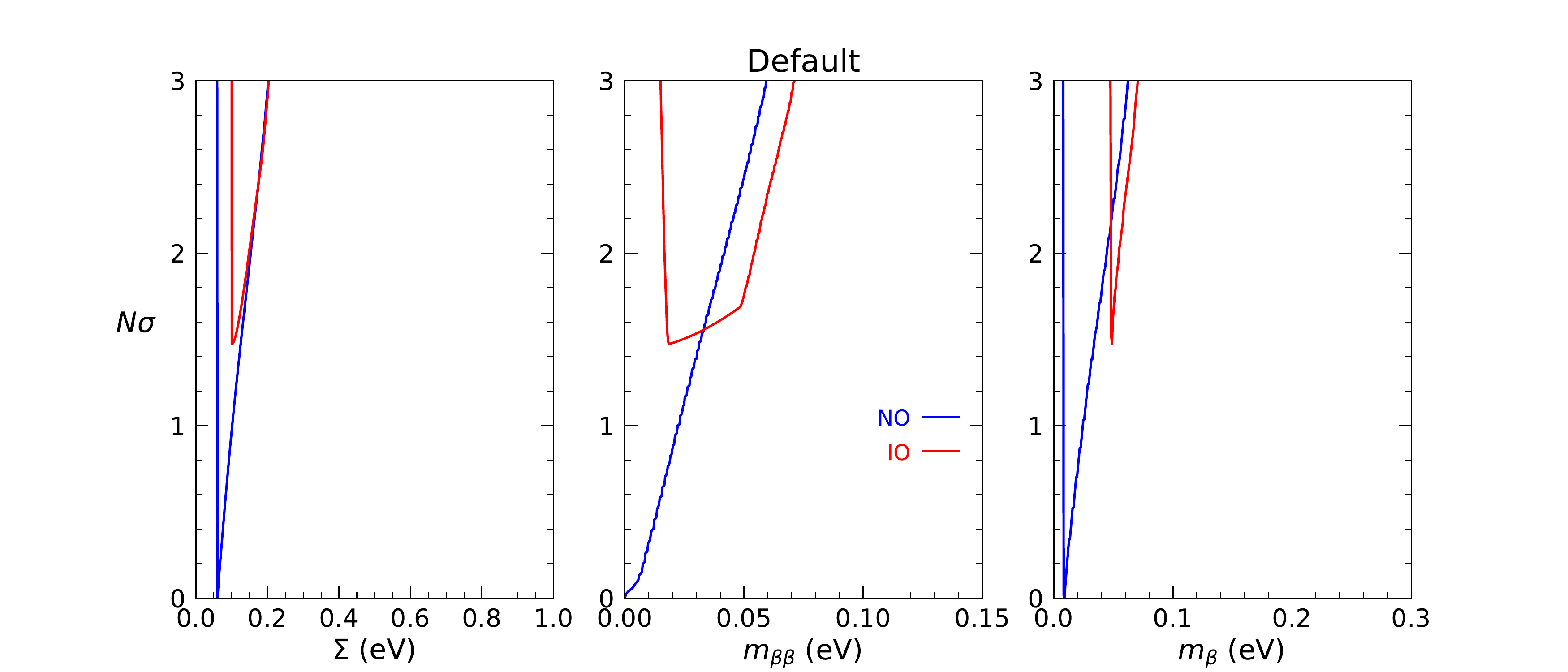}
\caption{\label{Fig_14}
\footnotesize 
$N_\sigma$ bounds on the single nonoscillation parameters 
$\Sigma$ (left), $m_{\beta\beta}$ (center) and $m_\beta$ (right), assuming default cosmological inputs. 
The combination of nonoscillation data induces the offset
between the absolute minima in IO (red) and NO (blue).
}
\end{minipage}
\end{figure}

\begin{figure}[h!]
\begin{minipage}[c]{0.85\textwidth}
\includegraphics[width=0.95\textwidth]{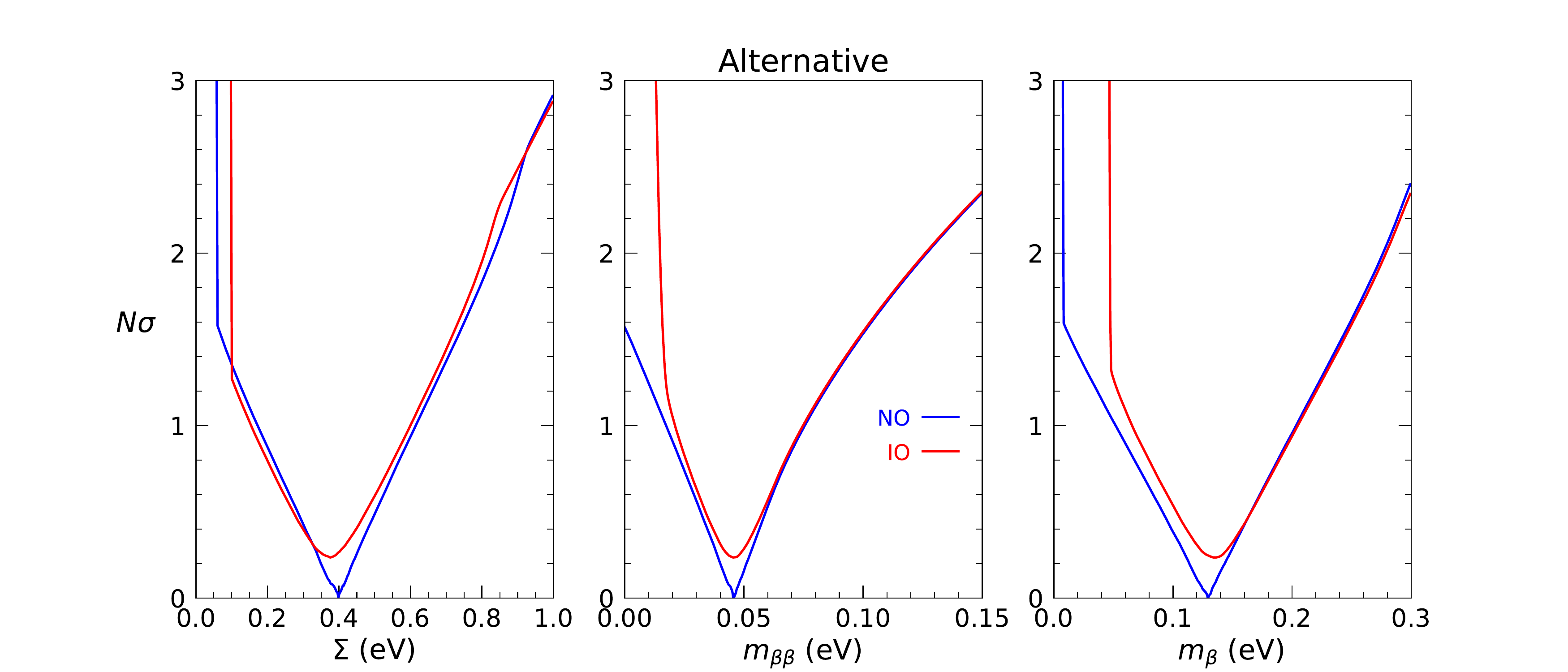}
\caption{\label{Fig_15}
\footnotesize 
As in Fig.~\ref{Fig_14}, but assuming alternative cosmological inputs. 
}
\end{minipage}
\end{figure}

\subsection{Results on single nonoscillation observables}

Figures~\ref{Fig_14} and \ref{Fig_15} show the projected bounds on single nonoscillation parameters for the 
default and alternative cases, respectively. In both cases we account for the NO--IO offset coming
from the combination of all nonoscillation data (i.e., the rightmost numbers in rows \#3 and \#12 of Table~\ref{Tab:Cosmo}
(that were omitted in the previous Figs.~\ref{Fig_12} and \ref{Fig_13}). A vertical rise of $N_\sigma$ occurs
when the lower physical limits are reached.
These two figures quantify 
previous considerations about the default and alternative options: the first exemplifies the case 
of strong upper bounds on $\Sigma$ from cosmology, accompanied by some sensitivity to the mass ordering, 
and by hard-to-probe phase spaces for $m_\beta$ and $m_{\beta\beta}$; the second represents the case of 
weaker upper bounds (and a possible signal) for $\Sigma$, with scarce sensitivity to the mass ordering but 
more optimistic expectations for $m_\beta$ and $m_{\beta\beta}$ signals. Together with 
\textcolor{black}{Fig.~\ref{Fig_03}},  
the above Figs.~\ref{Fig_14} and \ref{Fig_15} provide a neat summary 
of what we (do not) know in the standard $3\nu$ paradigm.%
\footnote{
\textcolor{black}{
At present, we stick to the viewpoint expressed in \cite{Fogli:2004as} and prefer to project away 
unobservable quantities, such as the lightest neutrino mass $m_0$ and the two Majorana phases $\eta_{1,2}$ (as defined in \cite{PDG1}). Of course,
when significant (and convergent) signals will emerge among the three observables
$(m_\beta,\,m_{\beta\beta},\,\Sigma)$, meaningful bounds on $m_0$ (and possibly weak hints on $\eta_{1,2}$) 
may also be derived.}
}

\section{Synthesis}

\begin{figure}[t!]
\begin{minipage}[c]{0.85\textwidth}
\includegraphics[width=0.8\textwidth]{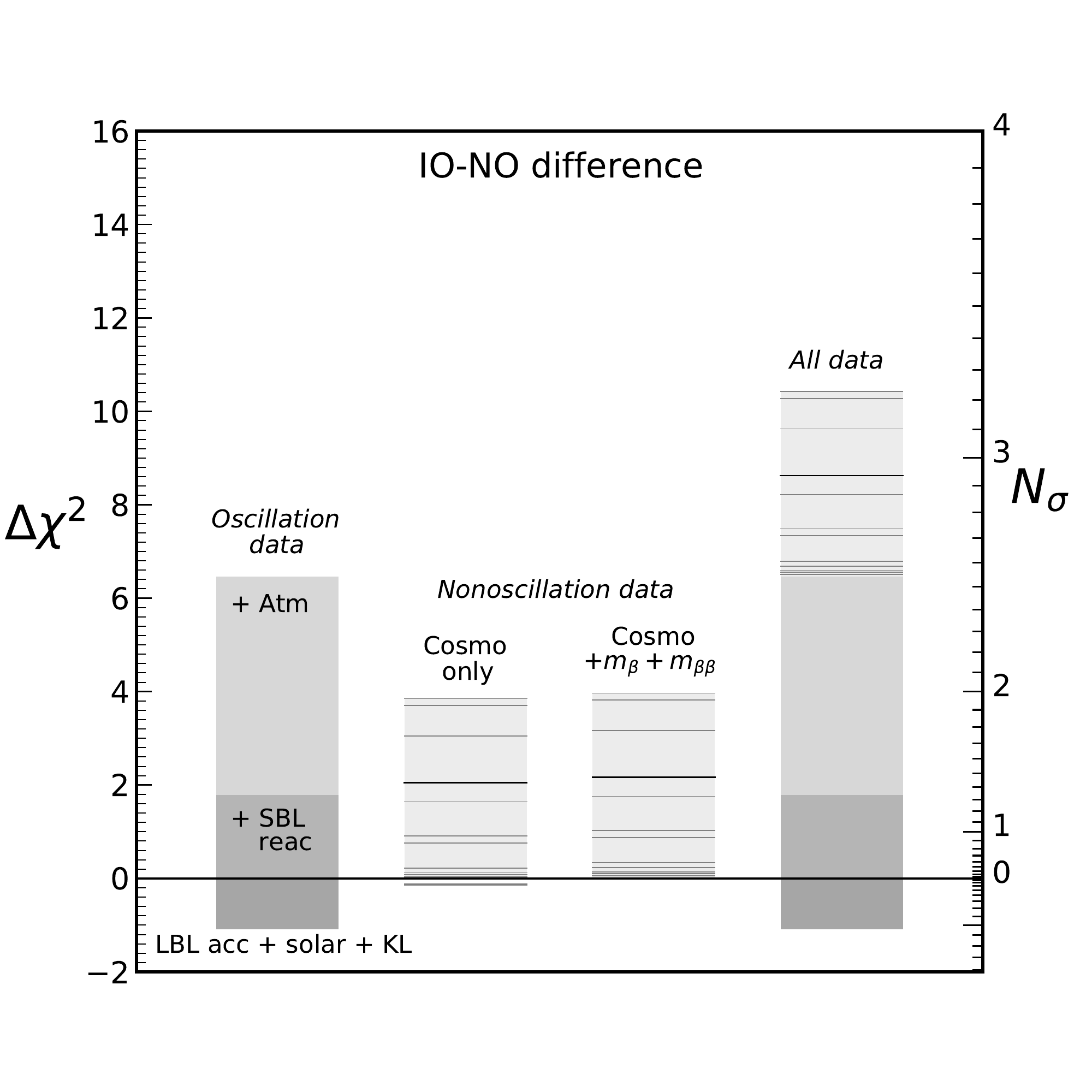}
\vspace*{-7mm}
\caption{\label{Fig_16}
\footnotesize 
Breakdown of contributions to the IO-NO $\chi^2$ difference from oscillation and nonoscillation data. The latter span the range 
of all the cosmological input variants reported in Table~\ref{Tab:Cosmo}, as indicated by the horizontal lines (the thick one 
corresponding to the  
default case).}
\end{minipage}
\end{figure}

We conclude our work by merging the information coming from the analysis of oscillation and nonoscillation data,
that have one important observable in common: the mass ordering.  Figure~\ref{Fig_16} shows a histogram with separate and combined contributions to the 
$\Delta\chi^2_{\mathrm{IO-NO}}$. The first bin adds up the contributions from oscillation data, starting from
the negative one in the combination of LBL accelerator, solar and KamLAND data, that becomes positive by adding SBL reactor data, and further increases with atmospheric data. The second bin shows the range spanned by all the cases considered in Table~\ref{Tab:Cosmo}, for the fit to cosmological data only.
 The third bin shows the slight change induced by adding 
current constraints on $m_\beta$ and $m_{\beta\beta}$, that provide an extra upward shift (see Tab.~\ref{Tab:Cosmo}).
The fourth bin adds up the contents of  the first and third bins, providing an overall indication
in favor of NO in the range $\sim 2.5$--$3.2\sigma$.  
 Although none of the single oscillation or nonoscillation data sets provides compelling evidence for 
 normal ordering yet, their current combination clearly favors this option at a global $\sim 3\sigma$ level. 

In conclusion, the main results of our global analysis can be essentially summarized in terms of bounds $N_\sigma=N_\sigma(p)$,
as shown in the following figures: 
Fig.~\ref{Fig_03} for the neutrino oscillation parameters 
$p=\delta m^2,\,
\textcolor{black}{|\Delta m^2|},
\,\theta_{12},\,\theta_{23},\,\theta_{13},\,\delta$; 
Figs.~\ref{Fig_14} and \ref{Fig_15} for the nonoscillation observables $p=m_\beta,\,m_{\beta\beta},\,\Sigma$,
in two representative cases for cosmological inputs; and Fig.~\ref{Fig_16} for the discrete mass ordering parameter, $p=\mathrm{sign}(\Delta m^2)=\mathrm{NO(+)/IO}(-)$.
Finishing the fabric of the $3\nu$ paradigm amounts to having convergent, 
narrow and linear $N_\sigma$ bounds, for one surviving mass
ordering, in terms of any continuous $3\nu$ oscillation parameter and nonoscillation observable $p$ (with 
the possible exception of $m_{\beta\beta}$, if neutrinos have a Dirac nature).
At present, this goal has been reached for $p=\delta m^2,\,
\textcolor{black}{|\Delta m^2|},
\,\theta_{12},\,\theta_{13}$
and, to some extent, for $p=\theta_{23}$ (up to an octant ambiguity). The current 
results for $p=\delta,\,m_\beta,\,m_{\beta\beta},\,\Sigma$ and NO/IO
may instead be considered as initial, shaky steps of a long march towards the characterization 
of the neutrino-antineutrino differences and of the absolute neutrino mass spectrum. 
On the way, we shall learn a lot about neutrino properties in many different contexts, 
clarify the origin of old and new data tensions, and possibly find obstacles
that, tearing away the fabric of the $3\nu$ paradigm,  may reveal hidden new physics.


\acknowledgments

We are grateful to M.\ Nakahata for informing us about the latest public release
of the Super-Kamiokande atmospheric neutrino (preliminary) analysis \cite{SKmap}.
We thank L.\ Pandola and M. Sisti for useful discussions about
$0\nu\beta\beta$ decay results.  This work is partly supported by the Italian Ministero dell'Universit\`a e Ricerca (MUR) through
the research grant number 2017W4HA7S ``NAT-NET: Neutrino and Astroparticle Theory Network'' under the program PRIN 2017,
and by the Istituto Nazionale di Fisica 
Nucleare (INFN) through the ``Theoretical Astroparticle Physics''  (TAsP) project. 
The work of F.C. is supported by the U.S.\ Department of
Energy under the award number DE-SC0020250.
E.D.V.\ acknowledges the support of the Addison-Wheeler Fellowship awarded by the Institute of Advanced Study at Durham University.


{}

\end{document}